\documentclass[11pt,a4paper]{article}
\usepackage{jheppub}
\usepackage{natbib}
\usepackage{graphicx}
\usepackage{cases}
\usepackage[dvipsnames]{xcolor}

\usepackage{mathtools}
\usepackage{hyperref}
\usepackage{float}
\usepackage{comment}
\usepackage{multirow}
\usepackage[overload]{empheq}
\usepackage[normalem]{ulem}
\usepackage{orcidlink}
\usepackage{rorlink}
\usepackage{bm}

\definecolor{Blu}{rgb}{0.,0.,1.}

\author[a]{Marco Cirelli$^{\ \orcidlink{0000-0002-4264-6323}}$}
\author[a]{\ \ Arpan Kar$^{\, \orcidlink{0000-0002-2993-3336}}$}
\author[b]{\ \ Halim Shaikh$^{\, \orcidlink{0009-0002-4465-4201}}$}
\affiliation[a]{Laboratoire de Physique Théorique et Hautes Énergies (\href{https://www.lpthe.jussieu.fr}{LPTHE})$^{\ \rorlink{https://ror.org/02mph9k76}}$,\\ CNRS $\&$ Sorbonne Université,
4 Place Jussieu, Paris, France}
\affiliation[b]{Fulbari, Karimpur, Nadia, West Bengal, India, 741158}

\emailAdd{marco.cirelli@gmail.com}
\emailAdd{arpankarphys@gmail.com}
\emailAdd{halimshaikh109@gmail.com}

\title{Indirect searches for realistic sub-GeV Dark Matter models}

\abstract{Indirect searches for Dark Matter (DM) particles with mass 
in the MeV -- GeV scale have received significant attention lately. 
Pair-annihilations of such DM particles in the Galaxy 
can give rise to (at the same time) MeV to GeV $\gamma$-rays via prompt emission, 
sub-GeV $e^\pm$ in cosmic-rays, as well as 
a broad photon spectrum ranging from $X$-rays to soft $\gamma$-rays, produced 
by the DM induced $e^\pm$ via inverse Compton scattering, bremsstrahlung 
and in-flight annihilation processes (collectively called `secondary emissions').  
We focus on two representative realistic sub-GeV DM models, namely, 
the vector-portal kinetic-mixing model and the higgs-portal model, 
and perform a detailed study of the indirect detection constraints 
from existing $X$-rays, $\gamma$-rays and cosmic-ray observations, 
based on all of the above-mentioned signals. We also estimate 
the future prospects from the upcoming MeV photon telescope {\sc Cosi}, 
including all possible types of prompt and secondary emission signals. We compare our results 
with the constraints and (or) projections from cosmological and terrestrial observations. 
We find that, for both the sub-GeV DM models, the current observations constrain the annihilation cross-section 
at the level of $\langle \sigma v \rangle \lesssim 10^{-27} {\rm cm}^3/{\rm s}$, or lower 
for some specific mass ranges or under optimistic assumptions. 
Moreover, new unconstrained DM parameter space can be probed at the upcoming instruments like {\sc Cosi}, thanks to the inclusion of secondary photons which in many cases provide the dominant signal.
}

\begin{document}

\maketitle

\section{Introduction}
\label{sec:introduction}
The idea that Dark Matter (DM) could be composed of a light particle, where light refers to a mass $m_{\rm DM}$ in the range of approximately 1 MeV to a few GeVs, has recently gathered significant attention. This interest stems partly from the absence of compelling evidence for the long-predicted weak-scale DM in current experiments, and partly from the rise of well-motivated theoretical models that predict such sub-GeV DM candidates (see for instance \cite{Knapen:2017xzo, Boehm:2002yz, Boehm:2003hm, Fayet:2007ua, Boehm:2003bt, Feng:2008ya, Hochberg:2014dra, Hochberg:2014kqa,Bertuzzo:2017lwt, Darme:2017glc} for a selection of recent works).

\medskip

The indirect detection of sub-GeV DM (i.e.~the searches for the products of annihilations --or decays-- of such particles in astrophysical environments) is a particularly promising, and somewhat still under-exploited, direction. 

From the observational point of view, the signals that can be looked for are essentially of three kinds\footnote{We focus the discussion, from now on, on annihilating sub-GeV DM, since this is the natural choice for the models we consider later. But our general analysis can be easily extended to the case of decaying sub-GeV DM as well.}: {\em (1)} MeV to GeV gamma rays emitted as prompt radiation from the DM annihilations, {\em (2)} sub-GeV electrons and positrons (collectively denoted in the following as $e^\pm$), emitted from the same annihilation processes, {\em (3)} soft gamma rays or keV $X$-rays emitted as secondary radiation from these $e^\pm$ in the astrophysical environment. 

From the phenomenological point of view, the analysis of these signals can be carried out in essentially two ways: {\em (a)} a model-independent one, i.e.~considering the annihilation of DM particles into representative channels such as $\gamma \gamma$, $e^+e^-$, $\mu^+\mu^-$ or $\pi^+\pi^-$, without committing to any specific particle physics realization; 
{\em (b)} a model-specific one, i.e.~assuming a fixed model of DM and a determined annihilation process into realistic final states, consisting of light hadronic resonances such as $\pi, \eta, K, \rho, \omega, \phi \ldots$.
The former approach has the advantage of being general and flexible. 
The latter, however, is instead particularly well motivated for sub-GeV DM: at these low energies, which are comparable to the QCD confining scale, the light mesons cited above {\em are} the relevant degrees of freedom in the final state of the annihilation (this is in contrast  to heavier DM models --above several GeVs-- in which the annihilation into quarks and gauge bosons describes well the process).

\medskip 

A number of works in the recent literature have tackled the indirect detection of sub-GeV DM along these observational and phenomenological axes. For the sake of setting the stage of the context of our study, we list here a selection of these works, making explicit  references to the signals {\em (1)}, {\em (2)} or {\em (3)} and to the approaches {\em (a)} or {\em (b)} above. 

Ref.~\cite{Essig:2013goa} studied the bounds from prompt photons (signal {\em (1)} above) measured by several MeV experiments, for a few generic channels (approach {\em (a)}).
Later on, ref.~\cite{Boddy:2015efa,Gonzalez-Morales:2017jkx,Caputo:2022dkz, ODonnell:2024aaw} considered the future prospects for the same signals, as did ref.~\cite{Bartels:2017dpb}, adding also secondary radiation {\em (3)}.
Refs.~\cite{Boudaud:2016mos, Boudaud:2018oya} derived bounds from {\sc Voyager-1} data on $e^\pm$ (signal {\em (2)} above), from generic annihilation channels {\em (a)}.
Ref.~\cite{Laha:2020ivk} considered the prompt monochromatic photons {\em (1)} from the annihilation in the $\gamma\gamma$ channel {\em (a)}, using  {\sc Integral} data. 
Refs.~\cite{Cirelli:2020bpc,Cirelli:2023tnx,Balaji:2025afr} computed the constraints from prompt {\em (1)} and secondary {\em (3)} radiation, in a model-independent {\em (a)} approach, using $X$-ray data from {\sc Integral}, {\sc NuStar}, {\sc Xmm-Newton}, {\sc eRosita} and other experiments. Ref.~\cite{DelaTorreLuque:2023olp} also included $e^\pm$ {\sc Voyager-1} data {\em (2)}.
Ref.~\cite{Coogan:2021sjs, Coogan:2021rez} worked out the current and projected limits from prompt photons {\em (1)}, as well as from the Cosmic Microwave Background (CMB), for higgs-portal and dark-photon-portal models, within approach {\em (b)} above. Similarly, ref.~\cite{Bernal:2025szh} considered prompt photons {\em (1)} in an ($L_\mu-L_\tau$) vector-mediator model, and ref.~\cite{Tang:2025vqf} in leptophilic DM models. Ref.~\cite{Watanabe:2025pvc} considered prompt photons {\em (1)} and $e^\pm$ {\em (2)} in specific effective models, and ref.~\cite{Nguyen:2024kwy} did the same for decaying DM. Ref.~\cite{Nguyen:2025tkl} adds secondary radiation {\em (3)}, always for decaying DM. 
Other recent studies on the broad phenomenology of sub-GeV detection 
include \cite{Balan:2024cmq,Cheek:2025nul, Lu:2024xxb,Linden:2024fby,Saha:2025wgg}. 

What this quick survey shows is that a study following the theoretically motivated approach {\em (b)} for annihilating DM, and considering all the {\em (1)}, {\em (2)} and {\em (3)} signal classes is currently missing.

\medskip 

The main focus of this work is therefore the following: focusing on representative realistic sub-GeV DM models, we perform a detailed study of the indirect detection constraints based on the existing photon and cosmic-ray observations, and we evaluate the future prospects from upcoming MeV photon telescopes like {\sc Cosi}, by including all possible types of prompt and secondary emission signals related to DM annihilations in the Galaxy. 

\medskip

The rest of this paper is organized as follows. 
In section \ref{sec:DM_models} we introduce the two classes of 
sub-GeV DM models that we consider in the following.
In section \ref{sec:photon_signal} we discuss the primary and secondary 
photon signals produced in the Galaxy from the annihilation of sub-GeV DM particles 
under such models. In section \ref{sec:cosmic_ray_signal} we discuss 
the corresponding cosmic-ray $e^\pm$ signal from DM annihilation. 
In section \ref{sec:obs_data_analysis} we discuss different observed data and describe the 
analysis methods. Section \ref{sec:MeV_telescope} contains a discussion on the 
upcoming MeV telescope {\sc Cosi}. In section \ref{sec:results_and_discussion} we presents 
our results and the corresponding discussions as well as the comparison 
with other constraints and projections. 
Finally, in section \ref{sec:Conclusions} we summarize our results and draw our conclusions. 
Appendix \ref{sec:Powers_secondary} offers some complementary technical details 
on secondary emissions, while appendix \ref{sec:DM_profiles} 
evaluates the impact of choosing different DM profiles.

\section{Models of sub-GeV Dark Matter}
\label{sec:DM_models}

In this section we introduce the two well-motivated sub-GeV particle DM models that we will consider in the following: a model featuring a vector mediator kinetically mixing with the  Standard Model (SM) photon (section \ref{sec:Vector_model}), and a scalar mediator mixing with the SM higgs (section \ref{sec:Scalar_model}). They are representative of generic classes: the vector-portal class~\cite{Fitzpatrick:2020vba, Abdughani:2021oit, Bernal:2025szh, Mohlabeng:2024itu} and the scalar-portal class~\cite{Krnjaic:2015mbs, Abdughani:2021oit, Tang:2025vqf, Arcadi:2019lka}, characterized by the fact that the DM particle interacts with the SM particles through a vector or a scalar mediator, respectively. And they correspond to two kinds of `portal' interactions routinely considered in the literature, see e.g.~\cite{Alexander:2016aln,LDMX:2018cma,Cirelli:2024ssz}.
In both cases we consider the DM to be a Dirac fermionic particle for definiteness, and we take both the DM and the mediator to be uncharged under the SM gauge group. 
We also assume symmetry between DM particles and antiparticles. 

\medskip

These sub-GeV DM models are implemented in the publicly-available code 
\texttt{HAZMA} \cite{Coogan:2019qpu, Coogan:2021sjs}   
and in its updated version \texttt{HAZMA2}~\cite{Coogan:2022cdd}. 
The codes provides accurate spectra of photons, electrons and positrons, 
and neutrinos produced in the pair-annihilation of DM particles. 
The spectrum of photons (produced via prompt emission) 
leads to the primary photon signals for indirect detection. The $e^\pm$ spectra, 
on the other hand, give rise to an analogous primary charged 
cosmic-ray signal but also to various secondary photon signals 
(which in many cases become stronger than the primary). 
These different signals will be discussed in 
sections.~\ref{sec:photon_signal} and \ref{sec:cosmic_ray_signal}. 

As mentioned in the introduction, in realistic sub-GeV DM models, where the annihilation typically occurs 
into more than one final state, the relevant final-state degrees of freedom are 
combinations of light hadronic 
resonances \cite{Coogan:2019qpu, Coogan:2022cdd}.\footnote{Note that the 
annihilation spectra obtained from the 
widely-used tool \texttt{PYTHIA}~\cite{Sjostrand:2007gs,Cirelli:2010xx} 
are reliable at center-of-mass energy above a few GeV, 
but not at lower energies, where strongly-interacting 
particles form hadronic bound states and can no longer be 
effectively described by parton showers, fragmentation and decay.}
\texttt{HAZMA}, using  chiral perturbation theory, computes the corresponding 
annihilation cross-sections (or the branching fractions of the total cross-section) 
as well as the final-state spectra up to a DM mass of 250 MeV. 
The updated version \texttt{HAZMA2}~\cite{Coogan:2022cdd} extends the range and 
provides the same outputs up to a center-of-mass energy of the 
DM annihilation of about 3 GeV (i.e. DM mass of 1.5 GeV). 
This is achieved in \texttt{HAZMA2} by bridging the regimes of validity 
of \texttt{HAZMA1} with that of another state of the art tool,  
\texttt{HERWIG4DM}~\cite{Plehn:2019jeo, Reimitz:2021wcq}, which is based on 
vector meson dominance and measured form-factors, and is accurate well up to a few GeVs. 
Altogether, these tools allow to compute the indirect detection spectra 
in the whole mass range of out interest, between 1 MeV and 1 GeV. 

\medskip

We now move to describe more precisely the two sub-GeV DM model classes analyzed in this work. See also \cite{Coogan:2019qpu, Coogan:2021rez, Coogan:2021sjs, Coogan:2022cdd} for the further detailed discussions. 

\subsection{Vector portal: dark photon or kinetic-mixing model}
\label{sec:Vector_model}
We consider a simplified model featuring a DM particle (a Dirac fermion for definiteness, 
as mentioned earlier) charged under 
a new $U(1)_D$ dark gauge group, whose vector boson $V$ mixes with the SM photon. 
The model is fully defined by the following parameters: 
the mass $m_{\rm DM}$ of the DM particle, the mass $m_V$ of the vector mediator $V$, the coupling of the vector mediator to DM ($g_{V{\rm DM}}$) and 
the kinetic mixing parameter $\epsilon$ between the vector mediator and the SM photon. 
The couplings of $V$ to the SM fermions are inherited from those of the photon and, explicitly, are given by $g_{Vf} = \epsilon e Q_f$, 
where $Q_f$ is the electric charge of the fermion $f$ and $e$ is the electron's charge. 
We assume the mediator's mass to be $m_V = 3 \, m_{\rm DM}$, 
for which most of the terrestrial constraints and future 
projections on this model (for sub-GeV DM masses) 
are usually available; see, for example, 
\cite{Alexander:2016aln, LDMX:2018cma, 10.1093/ptep/ptz106, Krnjaic:2022ozp, Antel:2023hkf}. 
This choice of the mediator's mass corresponds to the scenario dominated by  
`Direct Annihilation', where the annihilation of DM particles proceeds via 
${\rm DM} \, {\rm DM} \rightarrow V^{*} \rightarrow f\bar{f}$ 
to SM fermions through the virtual mediator. The total pair-annihilation 
cross-section of DM in this case is~\cite{Alexander:2016aln,LDMX:2018cma}:
\begin{equation}
\langle \sigma v \rangle \propto \epsilon^2 \, \alpha_D \, \frac{m^2_{\rm DM}}{m^4_V} 
= \frac{y}{m^2_{\rm DM}} \, , \hspace{8mm} {\rm with} \hspace{4mm} y = \epsilon^2 \, \alpha_D \, \left(\frac{m_{\rm DM}}{m_V}\right)^4 \, , 
\label{eq:sv_kineticMixing}
\end{equation}
where $\alpha_D = g^2_{V{\rm DM}} / 4\pi$. 
Note that here the DM annihilation proceeds through $s$-wave.

\medskip

\begin{figure*}[t]
\centering
\includegraphics[width=0.49\textwidth]{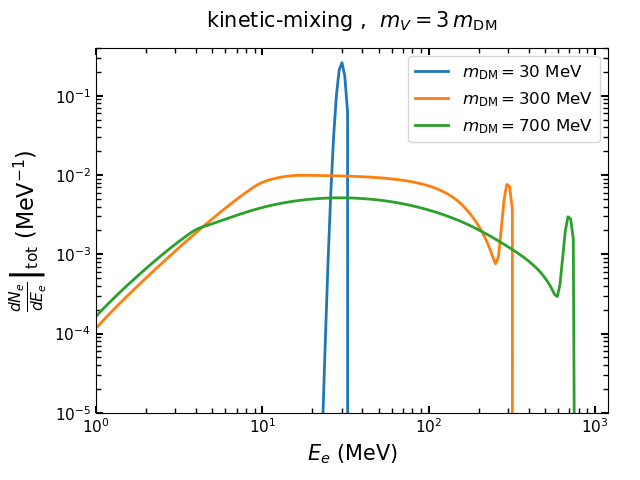}
\hspace{-2mm}
\includegraphics[width=0.49\textwidth]{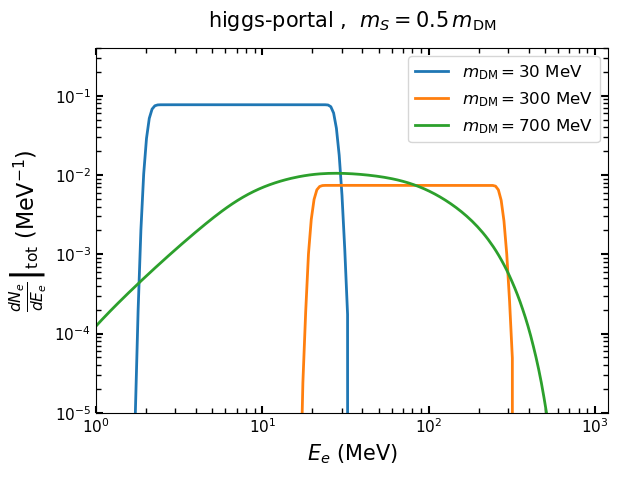}
\caption{\em Total electron or positron {\bfseries spectra} $\left.\frac{dN_e}{dE_e}\right\vert_{\rm tot}$ 
as a function of $E_e$ for different values of 
$m_{\rm DM}$, considering the kinetic-mixing model with $m_V = 3\,m_{\rm DM}$ (left panel) 
and higgs-portal model with $m_S = 0.5\,m_{\rm DM}$ (right panel). 
In each case, we add a 5\% resolution in the $e^\pm$ energy.}
\label{fig:dNdE_total}
\end{figure*}

As discussed above, we obtain the DM-induced annihilation spectra of $e^\pm$ and photons using \texttt{HAZMA}. 
A few examples of the $e^\pm$ spectrum $\left.\frac{dN_e}{dE_e}\right\vert_{\rm tot}$ 
for different DM masses ($m_{\rm DM} = 30, 300$ and 700 MeV) 
are shown in the left panel of fig.~\ref{fig:dNdE_total}. 
Here $\left.\frac{dN_e}{dE_e}\right\vert_{\rm tot} = \sum_F B_F \, \frac{dN^F_e}{dE_e}$ 
is the total spectrum as a sum of the contribution from each (kinematically possible) 
annihilation channel $F$ with the corresponding branching fraction $B_F$. 
The spectra consist of two main components. 
A sharp peak component, coming from the mediator ($V^*$) decaying into $e^+ \, e^-$, 
dominates for lower $m_{\rm DM}$'s and at higher energies for larger $m_{\rm DM}$'s. 
There is an additional smooth component (relevant for comparatively larger $m_{\rm DM}$'s), 
coming from mediator decays into muons and combinations of light mesons, 
which dominates the low energy part of the spectrum. See ref.~\cite{Coogan:2022cdd} 
for details. Note that in our parameter setup ($m_V = 3 \, m_{\rm DM}$) 
the spectra $\left.\frac{dN}{dE}\right\vert_{\rm tot}$ 
depend only on $m_{\rm DM}$ (or $m_V$), but not 
on the choice of $g_{V{\rm DM}}$ or $\epsilon$~\footnote{{In general, 
$\left.\frac{dN}{dE}\right\vert_{\rm tot} = \sum_F B_F \frac{dN^F}{dE}(m_{\rm DM}) = 
\sum_F \frac{\langle \sigma v \rangle_F}{\langle \sigma v \rangle_{\rm tot}} \frac{dN^F}{dE}(m_{\rm DM})$. 
Here $\frac{dN^F}{dE}$ is the $\gamma$ or $e^{+}$ spectrum produced in the SM channel $F$ and 
depends only on the outgoing energy in the channel $F$ (i.e., on the DM mass) but not on the DM couplings. 
In the case with $m_V = 3 \, m_{\rm DM}$, the only kinematically allowed annihilation channels are 
DM DM $\to V^* \to f \bar f$, for which $\langle \sigma v \rangle_F \propto (\epsilon \, g_{V{\rm DM}})^2$ 
(see Eq.~\ref{eq:sv_kineticMixing}), and hence 
$\langle \sigma v \rangle_{\rm tot} = \sum_F \langle \sigma v \rangle_F \propto (\epsilon \, g_{V{\rm DM}})^2$. 
As a result, the spectra $\left.\frac{dN}{dE}\right\vert_{\rm tot}$ are independent on $g_{V{\rm DM}}$ or $\epsilon$.}}, 
since the latter parameters enter only in determining the strength of $V^*$ production and decay. 
In the opposite regime ($m_V < m_{\rm DM}$), the competing annihilation channel 
${\rm DM}\, {\rm DM} \to VV \to f \bar f f^\prime \bar f^\prime$ (`Secluded Annihilation') opens up, 
producing different spectra, and the interplay among the two processes 
(hence the total spectra) depends on all parameters. 

\subsection{Scalar portal: higgs-portal model}
\label{sec:Scalar_model}
We consider a simplified model featuring a scalar mediator $S$ that mixes with the SM higgs boson. 
Like in the case of the kinetic-mixing model, here, too, 
it is assumed that the scalar mediator interacts 
with the SM particles only via its mixing with the SM higgs. 
The model is fully defined by the following parameters: 
the mass $m_{\rm DM}$ of the DM particle, the mass $m_S$ of the scalar mediator $S$, the coupling of the mediator to DM ($g_{S{\rm DM}}$) and the 
mixing angle $\vartheta$ between the scalar mediator and SM higgs. 
The couplings of the mediator to the SM fermions, gluons
and photon are inherited from those of the SM higgs. Namely, they are the same as those of the SM higgs except that they are suppressed by a factor  
$\sin\vartheta$. 
We assume the scalar mediator mass to be $m_S = m_{\rm DM} / 2$, i.e.~lower than the DM mass. 
In the other case, where the mediator is heavier than the DM, 
different terrestrial search constraints are 
expected to dominate over those coming from 
indirect detection searches (see, e.g., \cite{Coogan:2021sjs, Coogan:2021rez}).
In the case that we consider, the annihilation of DM 
into SM particles is dominated by the process ${\rm DM} \, {\rm DM} \rightarrow S S$ 
followed by the decay of $S$ into SM particles. 
The corresponding total pair-annihilation cross-section is~\cite{Krnjaic:2015mbs}: 
\begin{equation}
\langle \sigma v \rangle = \frac{3 \, g^4_{S{\rm DM}} \, v^2_{\rm rel}}{128 \, \pi \, m^2_{\rm DM}} \, , 
\label{eq:sv_HiggsPortal}
\end{equation}
with $v_{\rm rel}$ being the relative velocity of the two annihilating DM particles. 
Note that here $\sin\vartheta$ plays no role (so the thermal freeze-out is 
compatible with a wide range of values of $\vartheta$) 
and the DM annihilation proceeds through $p$-wave. 
For this model we will present (in sec.~\ref{sec:results_and_discussion}) 
all constraints on $\langle \sigma v \rangle$ that corresponds to 
our galaxy (with $v_{rel} \simeq 220$ km/s). 

\medskip

A few examples of the total $e^\pm$ spectrum $\left(\left.\frac{dN_e}{dE_e}\right\vert_{\rm tot}\right)$ 
in this model are shown in the right panel of fig.~\ref{fig:dNdE_total} 
for different DM masses ($m_{\rm DM} = 30, 300$ and 700 MeV). 
The spectra consist of a box component coming
from the {\em on-shell} mediator $S$ decaying into $e^+ \, e^-$ 
(which dominates at lower DM masses for $m_{\rm DM}$ 
above the electron threshold, 2 MeV), 
and a smooth component coming from $S$ decaying into muons and pions 
(which is dominating at higher DM masses). 
See \cite{Coogan:2019qpu, Coogan:2021rez, Coogan:2021sjs} for details. 
Note that in our parameter regime ($m_S = m_{\rm DM} / 2$) 
the spectra $\left.\frac{dN}{dE}\right\vert_{\rm tot}$ for $e^\pm$ or photon 
depends only on $m_{\rm DM}$ (or $m_S$), but not on the choice of $g_{S{\rm DM}}$ or $\sin\vartheta$ 
(at least for not-too-extreme values of these parameters, otherwise the competing annihilation channel 
DM DM $\to S \to f \bar f$ becomes relevant and, since it depends $g_{S{\rm DM}}$ and $\sin\vartheta$, the total spectra are modified)~\footnote{{As mentioned in sec.~\ref{sec:Vector_model}, 
$\left.\frac{dN}{dE}\right\vert_{\rm tot} = 
\sum_F \frac{\langle \sigma v \rangle_F}{\langle \sigma v \rangle_{\rm tot}} \frac{dN^F}{dE}(m_{\rm DM})$. 
As long as only ${\rm DM}\, {\rm DM} \to VV \to f \bar f f^\prime \bar f^\prime$ annihilation dominates, 
$\langle \sigma v \rangle_F$ and thus $\langle \sigma v \rangle_{\rm tot}$ $(= \sum_F \langle \sigma v \rangle_F)$ 
are proportional to $g^4_{S{\rm DM}}$ (Eq.~\ref{eq:sv_HiggsPortal}). 
Hence, $\left.\frac{dN}{dE}\right\vert_{\rm tot}$ is independent on $g_{S{\rm DM}}$ or $\sin\vartheta$.}}.

\bigskip

In the next two sections we discuss various possible indirect detection 
signals that can arise due to the 
annihilation of sub-GeV DM particles under a realistic model. 
These discussions on signals will be focused on the 
kinetic-mixing DM model discussed in the previous sub-section, for definiteness, 
but they are easily generalized to the other DM models.

\section{Photon signals from sub-GeV DM annihilation in the inner Galaxy}
\label{sec:photon_signal}

In this section we discuss the photon signals from 
the annihilation of sub-GeV DM in the inner Galaxy.  
We assume that the density distribution of DM in the galactic halo is 
described by the NFW profile~\cite{Navarro:1995iw}: 
\begin{equation}
\rho_{\rm DM}(r) = \frac{\rho_0}{\left(\frac{r}{r_s}\right) 
\left(1 + \frac{r}{r_s}\right)^2} \, ,
\label{eq:rho_profile}
\end{equation}
with $r$ as the radial distance from the galactic center (GC). 
For the parameters $\rho_0$ and $r_s$ {we consider the values 
$\rho_0 = 0.839 \, {\rm GeV\,cm^{-3}}$ and $r_s = 11$ kpc, which correspond to} 
the fitted NFW profile parameters (the central values) 
obtained in \cite{2019JCAP...10..037D} for the baryonic model B2. 
However, for comparison we will also consider
other choices for the DM profile, see App.~\ref{sec:DM_profiles}. 

The photon energy range considered in this work is 
$10 \, {\rm keV} \lesssim E_\gamma \lesssim 1 \, {\rm GeV}$ which 
includes the existing observations by {\sc Integral}, {\sc Comptel}, {\sc Egret} 
and {\sc Fermi-Lat} (partially) as well as the future observation by the 
MeV telescope {\sc Cosi}. As for the region of observation (ROI), 
we consider areas of different sizes (centered on the GC) 
corresponding to different experimental observations, i.e., 
$|b| < 15^{\circ} , |l| < 30^{\circ}$ for {\sc Integral}, 
$|b| < 20^{\circ} , |l| < 60^{\circ}$ for {\sc Comptel}, 
$|b| < 10^{\circ} , |l| < 60^{\circ}$ for {\sc Egret} and 
$\theta<10^{\circ}$ for {\sc Fermi} as well as for {\sc Cosi} observations. 
The coordinates $(b, l)$ denote the galactic latitude and longitude, 
which are related to the angular distance $\theta$ from the GC 
as $\cos \theta = \cos b \, \cos l$. 
The photon signal from such galactic regions 
over the energy range of interest gets contributions from 
both the prompt $\gamma$-ray emission as well as from 
various secondary emission processes. 
We discuss such processes below. 
As we will see, in several cases, depending on the 
DM mass and the observation energy ranges, secondary emission 
signals can dominate over the prompt one. 

\subsection{Primary signal: prompt $\gamma$-ray emission}
\label{sec:prompt}

The photon flux through the prompt emission 
(averaged over the observation region $\Delta \Omega$) is given by 
(see e.g.~\cite{Cirelli:2024ssz,Cirelli:2025qxx}: 
\begin{equation}
\frac{d\Phi_{\rm prompt}}{dE_\gamma} = \frac{\langle \sigma v \rangle}{8 \pi \, f_\chi \, m^2_{\rm DM}} \, 
\left.\frac{dN_\gamma}{dE_\gamma}\right\vert_{\rm tot} \, 
\frac{J_{\Delta \Omega}}{\Delta \Omega} \, ,
\label{eq:prompt_flux}
\end{equation}
where $\langle \sigma v \rangle$ is the total velocity-averaged annihilation cross-section of DM particles 
and $\left.\frac{dN_\gamma}{dE_\gamma}\right\vert_{\rm tot}$ is the total photon energy spectrum produced 
per annihilation of DM particles of mass $m_{\rm DM}$ under a given DM model. 
The factor $f_\chi$ takes the values 1 or 2 depending on whether the DM particle is self-conjugate or not. 
Since in our considered models the DM is a Dirac fermion, $f_\chi = 2$. 
As discussed in sec.~\ref{sec:DM_models}, the spectrum $\left.\frac{dN_\gamma}{dE_\gamma}\right\vert_{\rm tot}$ 
for the DM mass range considered in this work is obtained 
from \texttt{HAZMA2} (with an added energy resolution of 5\%).
The annihilation $J$-factor for the ROI $\Delta \Omega$ is defined as usual as:
\begin{equation}
J_{\Delta \Omega} = \int_{\Delta \Omega} d\Omega \, \int_{\rm l.o.s.} ds \,\, \rho^2_{\rm DM} (r(s,\theta)) \, \, ,
\end{equation}
where $s$ runs along the line-of-sight ($l.o.s.$) 
(with respect to the observer) at an angular distance $\theta$ from the GC. 

\begin{figure*}[ht!]
\centering
\includegraphics[width=0.75\textwidth]{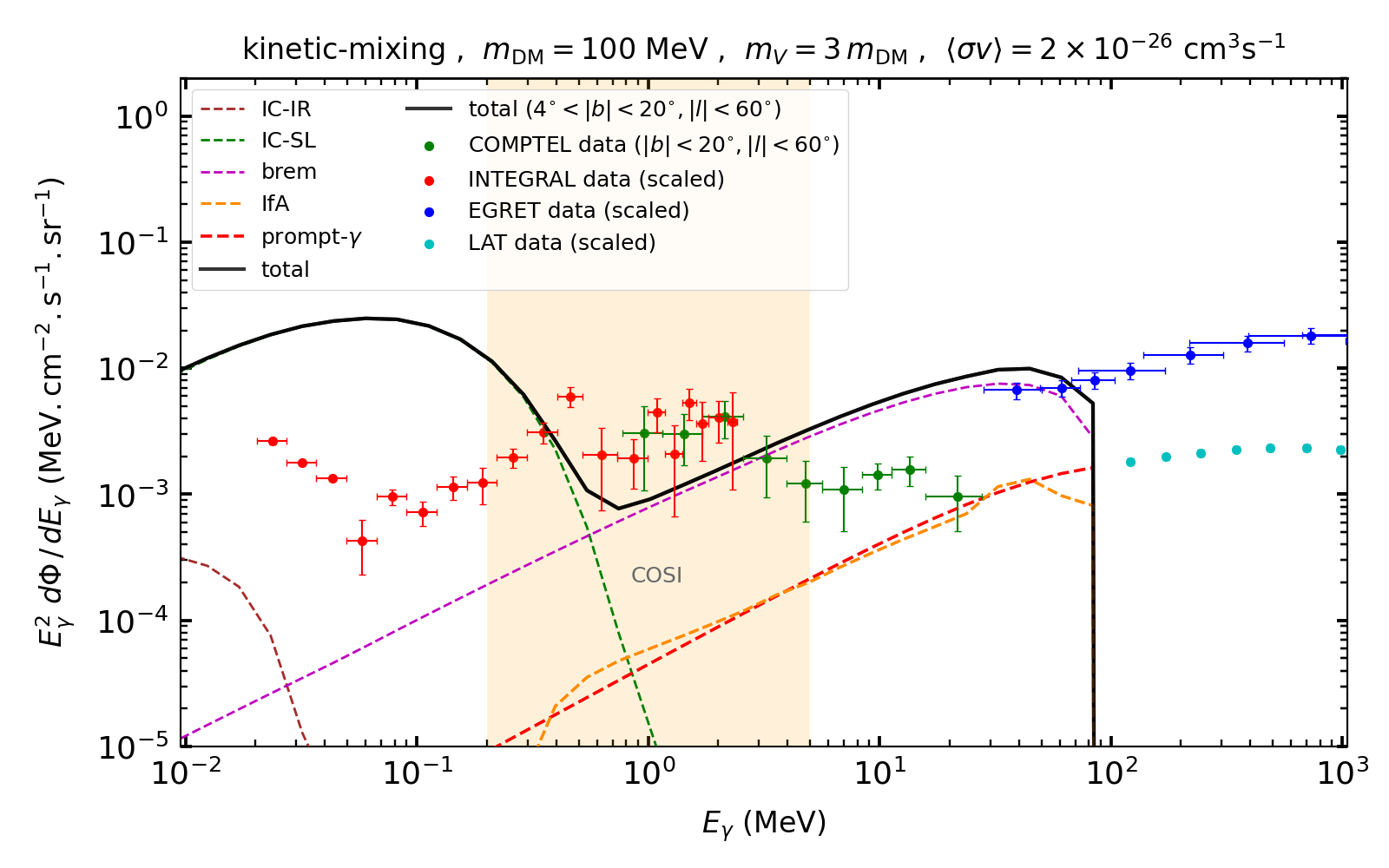}
\vspace{-4mm}
\caption{\em Illustration of the different components (prompt and secondaries) 
of the {\bfseries total photon flux} (solid black line) 
produced due to the annihilation 
of DM particles (of mass $m_{\rm DM} = 100$ MeV) in the Galaxy are shown 
for the considered photon energy range. 
The DM interaction is driven by the kinetic-mixing model 
with $m_V = 3 \, m_{\rm DM}$ and the total annihilation cross-section 
$\langle \sigma v \rangle = 2\times10^{-26}$ $\rm cm^3 s^{-1}$. 
The distribution of DM-induced $e^\pm$ (that give rise to the secondaries) 
in the Galaxy is obtained considering the approach discussed 
in sec.~\ref{sec:semi-analytic}. 
All DM signals (including the total one) are estimated 
from the region $4^{\circ} < |b| < 20^{\circ} , |l| < 60^{\circ}$, 
a region similar to the one covered by {\sc Comptel} observation, 
but masking $|b| \leq 4^{\circ}$. The red, green, blue and cyan data points are 
from the observations of {\sc Integral} ($|b| < 15^{\circ} , |l| < 30^{\circ}$), 
{\sc Comptel} ($|b| < 20^{\circ} , |l| < 60^{\circ}$), 
{\sc Egret} ($|b| < 10^{\circ} , |l| < 60^{\circ}$) 
and {\sc Fermi-Lat} ($\theta<10^{\circ}$), respectively 
(as discussed in sec.~\ref{sec:obs_data_analysis}). 
All data presented in the figure are rescaled to the observation region of {\sc Comptel}.  
The light orange band indicates the energy range covered by 
upcoming MeV telescope {\sc Cosi} (see sec.~\ref{sec:MeV_telescope}). 
}
\label{fig:DM_fluxes}
\end{figure*}

As an illustration, in fig.~\ref{fig:DM_fluxes} we show by the red dashed line 
the prompt $\gamma$-ray signal arising from the annihilation of 
galactic DM particles (of mass $m_{\rm DM} = 100$ MeV). 
The DM interaction with the SM is assumed to driven by the 
kinetic-mixing model with $m_V = 3 \, m_{\rm DM}$ and  
total annihilation cross-section 
$\langle \sigma v \rangle = 2\times10^{-26}$ $\rm cm^3 s^{-1}$. 
Here the photon signal (eq.~\ref{eq:prompt_flux}) 
is estimated from the region $4^{\circ} < |b| < 20^{\circ} , |l| < 60^{\circ}$, 
 similar to the region covered by the {\sc Comptel} observations, 
but masking $|b| \leq 4^{\circ}$ 
(as also considered in the next sub-section for the secondary emissions). 

\subsection{Secondary signals: Inverse Compton Scatterings, 
bremsstrahlung and In-flight annihilation}
\label{sec:secondaries}

The photon flux corresponding to the secondary emission 
(averaged over the observation region $\Delta \Omega$)
induced by DM annihilation in the Galaxy can be estimated as:
\begin{equation}
\frac{d\Phi_{\rm sec}}{dE_\gamma} = 
\frac{1}{\Delta \Omega} \int_{\Delta \Omega} d\Omega \, 
\left[\frac{1}{E_\gamma} \int_{\rm l.o.s.} ds \,\, \frac{j_{\rm sec} (E_\gamma, \Vec{x}(s,b,l))}{4\pi} \right] \, ,
\label{eq:secondary_flux}
\end{equation}
where $j_{\rm sec}$ is the photon emissivity 
(at a position $\Vec{x}$ in the Galaxy) 
corresponding to one of the secondary processes: 
(i) Inverse Compton Scatterings (ICS) on the interstellar radiation fields (ISRF) 
composed of CMB, infrared (IR) and starlight (SL), 
(ii) bremsstrahlung on interstellar gas particles and 
(iii) in-flight annihilation (IfA) of positrons.  
These emissivities can be expressed as (see e.g.~\cite{Cirelli:2024ssz}): 
\begin{eqnarray}
j_{\rm ICS} (E_\gamma, \Vec{x}(s,b,l)) &=& 2 \, \int^{m_{\rm DM}}_{m_e} dE_e \, \mathcal{P}_{\rm ICS} (E_\gamma, E_e, \Vec{x}) 
\,\, \frac{dn_{e}}{dE_e}(E_e,\Vec{x}) \, , 
\label{eq:j_ICS}
\\
j_{\rm brem} (E_\gamma, \Vec{x}(s,b,l)) &=& 2 \, \int^{m_{\rm DM}}_{m_e} dE_e \, \mathcal{P}_{\rm brem} (E_\gamma, E_e, \Vec{x}) 
\,\, \frac{dn_{e}}{dE_e}(E_e,\Vec{x}) \, ,
\label{eq:j_brem}
\\
j_{\rm IfA} (E_\gamma, \Vec{x}(s,b,l)) &=& \int^{m_{\rm DM}}_{m_e} dE_e \, \mathcal{P}_{\rm IfA} (E_\gamma, E_e, \Vec{x}) 
\,\, \frac{dn_{e}}{dE_e}(E_e,\Vec{x}) \, .
\label{eq:j_IfA}
\end{eqnarray}
Here $\mathcal{P}_{\rm ICS}$, $\mathcal{P}_{\rm brem}$ and $\mathcal{P}_{\rm IfA}$ 
denote the powers emitted (per photon energy) in the relevant secondary processes 
into photons with energy $E_\gamma$ by an $e^+$ or $e^-$ (having an energy $E_e$); 
they are given for reference in appendix~\ref{sec:Powers_secondary}. 
The function $\frac{dn_{e}}{dE_e}(E_e,\Vec{x})$ 
in eqs.~\eqref{eq:j_ICS}-\eqref{eq:j_IfA}
denotes the differential number density distribution of the $e^\pm$ 
arising due to the DM annihilation~\footnote{For the spatial integral in 
eq.~\eqref{eq:secondary_flux}, we cut the distribution $\frac{dn_e}{dE_e}(E_e,\Vec{x})$ 
in the radial direction at 
$R = R_{\rm Gal} = 20 \, {\rm kpc}$, and in the vertical direction  
at $|z| = L_{\rm Gal} = 4 \, {\rm kpc}$, 
assuming this to be the size of the zone that keeps the $e^\pm$ confined.}. 
Note that the factor of 2 in \eqref{eq:j_ICS} and \eqref{eq:j_brem} takes into account 
the contributions of both the population of positrons and electrons induced by the DM annihilation.  

The distribution $\frac{dn_{e}}{dE_e}$ results due to the injection 
of the $e^\pm$'s by DM annihilation in a given model. 
The rate of such injection can be quantified in terms of the source function as:
\begin{equation}
Q_e (E^S_e, r) = \frac{\langle \sigma v \rangle}{2 \, f_\chi \, m^2_{\rm DM}} \, 
\left.\frac{dN_e}{dE^S_e}\right\vert_{\rm tot} \, \rho^2_{\rm DM}(r) \, ,
\end{equation}
with $f_\chi = 2$ (for Dirac fermion DM). 
Here $\left.\frac{dN_e}{dE^S_e}\right\vert_{\rm tot}$ 
is the total energy spectrum of $e^\pm$ 
sourced by the annihilation of DM of mass $m_{\rm DM}$ in a given model. 
Such a spectrum is obtained using \texttt{HAZMA2} 
(with an added energy resolution of 5\%), 
as discussed in sec.~\ref{sec:DM_models}.

\subsubsection{Semi-analytic approach to obtain the $e^\pm$ distribution} 
\label{sec:semi-analytic}

The electrons and positrons, after being produced from the DM annihilation, 
propagate through the galactic medium undergoing several phenomena 
including spatial diffusion and radiative energy losses 
and give rise to a steady state distribution. 
We will be interested in observing the photon signals from 
the regions around the GC where the effect of energy losses 
are expected to dominate over other processes. 
In such a case, the spatial and energy 
distribution of the steady state $e^\pm$'s can be obtained 
in a semi-analytic way as~ (see e.g.~\cite{Cirelli:2009vg, Cirelli:2025qxx}): 
\begin{equation}
\frac{dn_e}{dE_e} (E_e, \Vec{x}) = \frac{1}{b_{\rm tot} (E_e, \Vec{x})} 
\int^{m_{\rm DM}}_{E_e} dE^S_e \, Q_e (E^S_e, r) \, ,
\label{eq:dnedE_analytic}
\end{equation}
where $b_{\rm tot} (E_e, \Vec{x})$ denotes the total energy loss rate of the 
$e^\pm$ in the galactic medium via various processes. 
These (in order of importance with increasing $e^\pm$ energy) are: 
Coulomb interactions with the interstellar gases, 
ionization of the same gases, bremsstrahlung on the same gases, 
synchrotron emission in the galactic magnetic field and 
ICS off the ambient photons in the ISRF. 
See \cite{Buch:2015iya} for the detailed expressions of these energy losses.
The maps for galactic gases, ISRF photons and magnetic field 
are obtained from \cite{Buch:2015iya} and are same 
as the ones used in \cite{Cirelli:2020bpc, Cirelli:2023tnx} as well as 
in \cite{Cirelli:2025qxx}.  

In this semi-analytic approach, 
in order to be on the conservative side, 
we estimate the DM-induced secondary (eq.~\eqref{eq:secondary_flux}) 
as well as the primary (eq.~\eqref{eq:prompt_flux}) photon signals 
from a region $|b| > 4^{\circ}$, i.e., applying a mask of $4^{\circ}$ 
above and below the Galactic Plane (GP). 
This process excludes most of the signal that can come from the GP. 

In fig.~\ref{fig:DM_fluxes} we show different secondary photon signals 
(corresponding to ICS off different ambient photons, 
bremsstrahlung and in-flight annihilation)
produced as a result of the annihilation of 
galactic DM particles (of mass $m_{\rm DM} = 100$ MeV) in the context of 
the kinetic-mixing model (with $m_V = 3 \, m_{\rm DM}$). 
The DM-induced $e^\pm$ distribution that gives rise to such signals are 
obtained using eq.~\eqref{eq:dnedE_analytic}. 
The secondary signals are estimated from the region 
$4^{\circ} < |b| < 20^{\circ} , |l| < 60^{\circ}$. 
Fig.~\ref{fig:DM_fluxes} shows that, in the energy range considered here, 
the secondary photons can in principle dominate the total DM signal. 

\subsubsection{Full propagation of $e^\pm$ in the Galaxy} 
\label{sec:full_propagation}

We also consider the full galactic propagation of the DM-induced $e^\pm$. 
This involves, apart from the radiative energy losses, 
various other processes, such as spatial diffusion, advection, 
convection and re-acceleration. The final distribution 
($\frac{dn_{e}}{dE_e}(E_e,\Vec{x})$) resulting after all such processes 
can be obtained by solving a differential equation representing the 
transport or propagation of $e^\pm$; 
see e.g.~eq.~2.1 of \cite{Evoli:2016xgn} and \cite{Evoli:2017vim}. 
We solve such an equation using the numerical 
package \texttt{DRAGON2}~\cite{Evoli:2016xgn, Evoli:2017vim},  
considering two Galactic propagation
models, named here as `prop.~a' and `prop.~b'. 
These models were used in \cite{Orlando:2017mvd} 
as baseline models for the study related to the secondary photon 
emissions from astrophysical origins towards the GC. 
Such galactic emissions were later used in \cite{Negro:2021urm} 
as fiducial photon backgrounds in the context of the upcoming 
MeV telescopes such as {\sc Cosi}. 

\underline{\bf Model `prop. a'}: It refers to the propagation model 
`DRELowV' from \cite{Orlando:2017mvd}, where the spatial diffusion is 
parameterized as a function of the propagating particle rigidity ($R$) as 
$D (R) = D_0 \, \beta_e \, (R / R_0)^\delta$, 
with $\beta_e$ as the dimensionless particle velocity. 
The values of $D_0$, $R_0$ and $\delta$ are 
$1.46\times10^{29}$ ${\rm cm^2 s^{-1}}$, 40 GV and 
$\sim0.33$, respectively. The Alfvén velocity  
(related to the re-acceleration of the particles) is $v_A = 8.9$ km/s. 
As shown in \cite{Orlando:2017mvd}, the secondary fluxes 
corresponding to this model are similar to that obtained 
with their other baseline model `PDDE'. 

\underline{\bf Model `prop. b'}: We also consider this propagation model 
which refers to the model `DRC' from \cite{Orlando:2017mvd}. 
In this model we have 
$D_0 = 4.3\times10^{28}$ ${\rm cm^2 s^{-1}}$, 
$R_0 = 4$ GV, $\delta \simeq 0.395$ and a larger Alfvén velocity 
$v_A = 28.6$ km/s. In addition, it includes a convective wind velocity 
in terms of its magnitude close to the GP, $v_c = 12.4$ km/s, 
and its gradient perpendicular to the GP, $dv_c/dz = 10.2$ km/s/kpc.  

\begin{figure*}[ht!]
\begin{minipage}{0.67\textwidth}
\includegraphics[width=\textwidth]{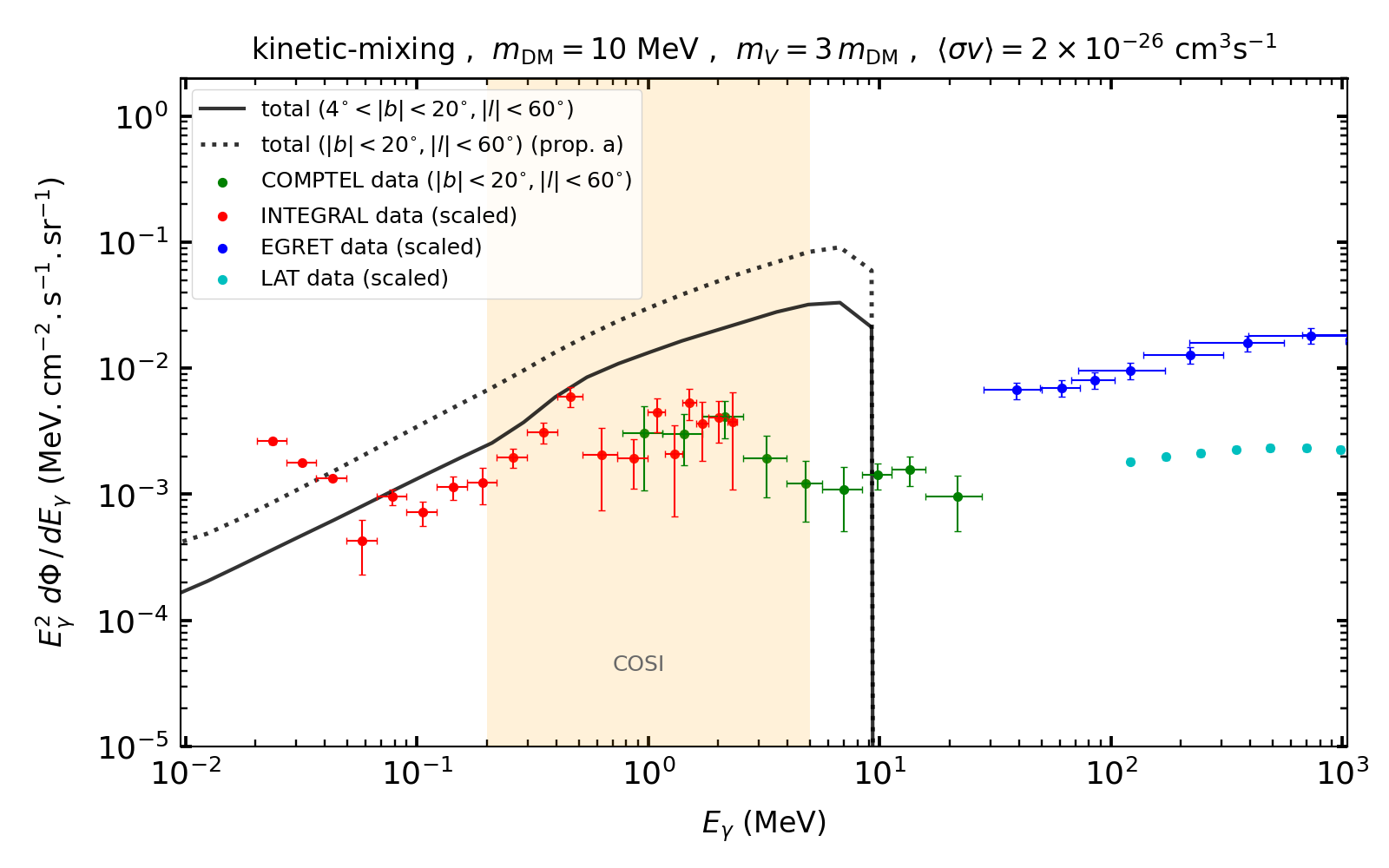}\\
\includegraphics[width=\textwidth]{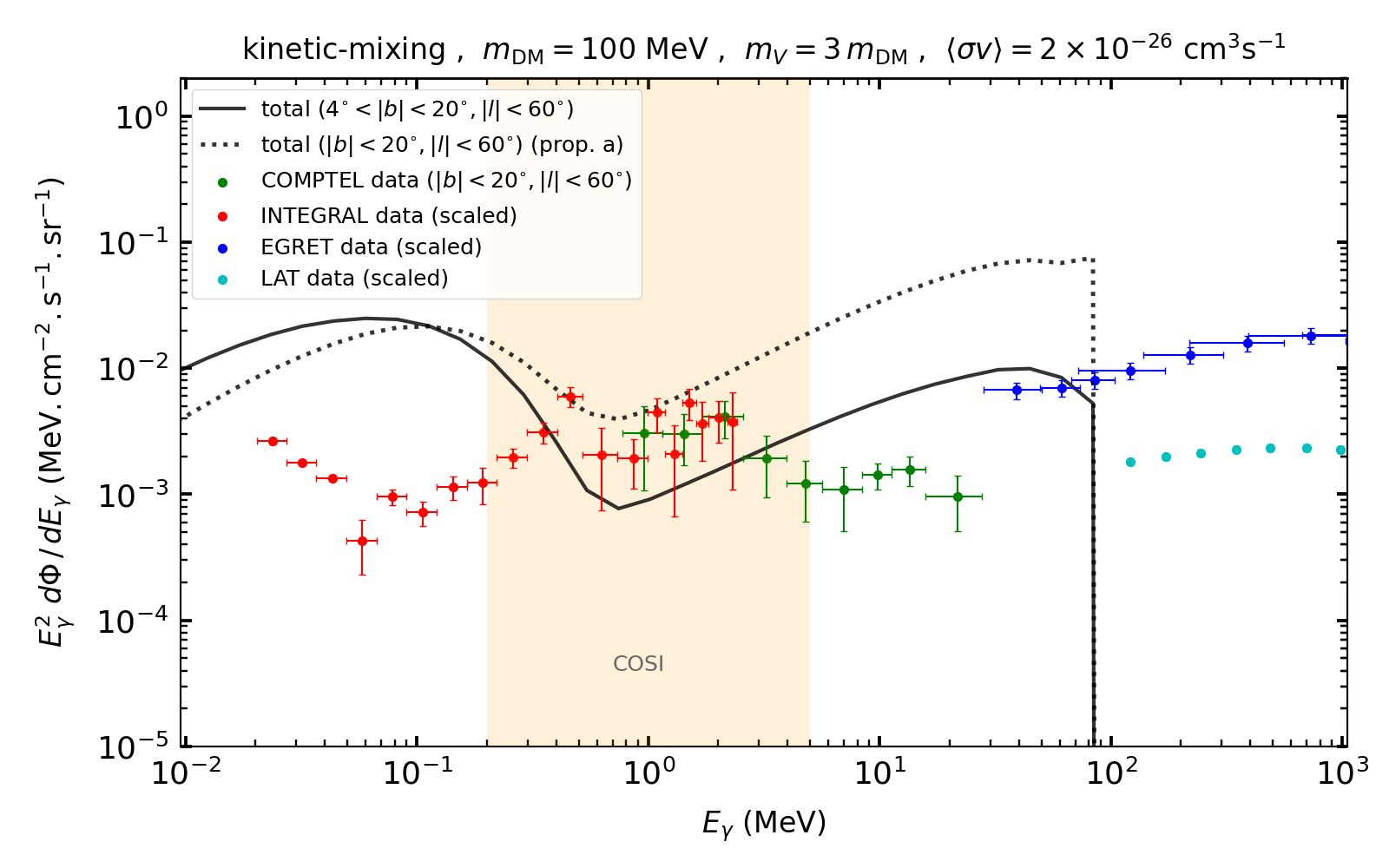}\\
\includegraphics[width=\textwidth]{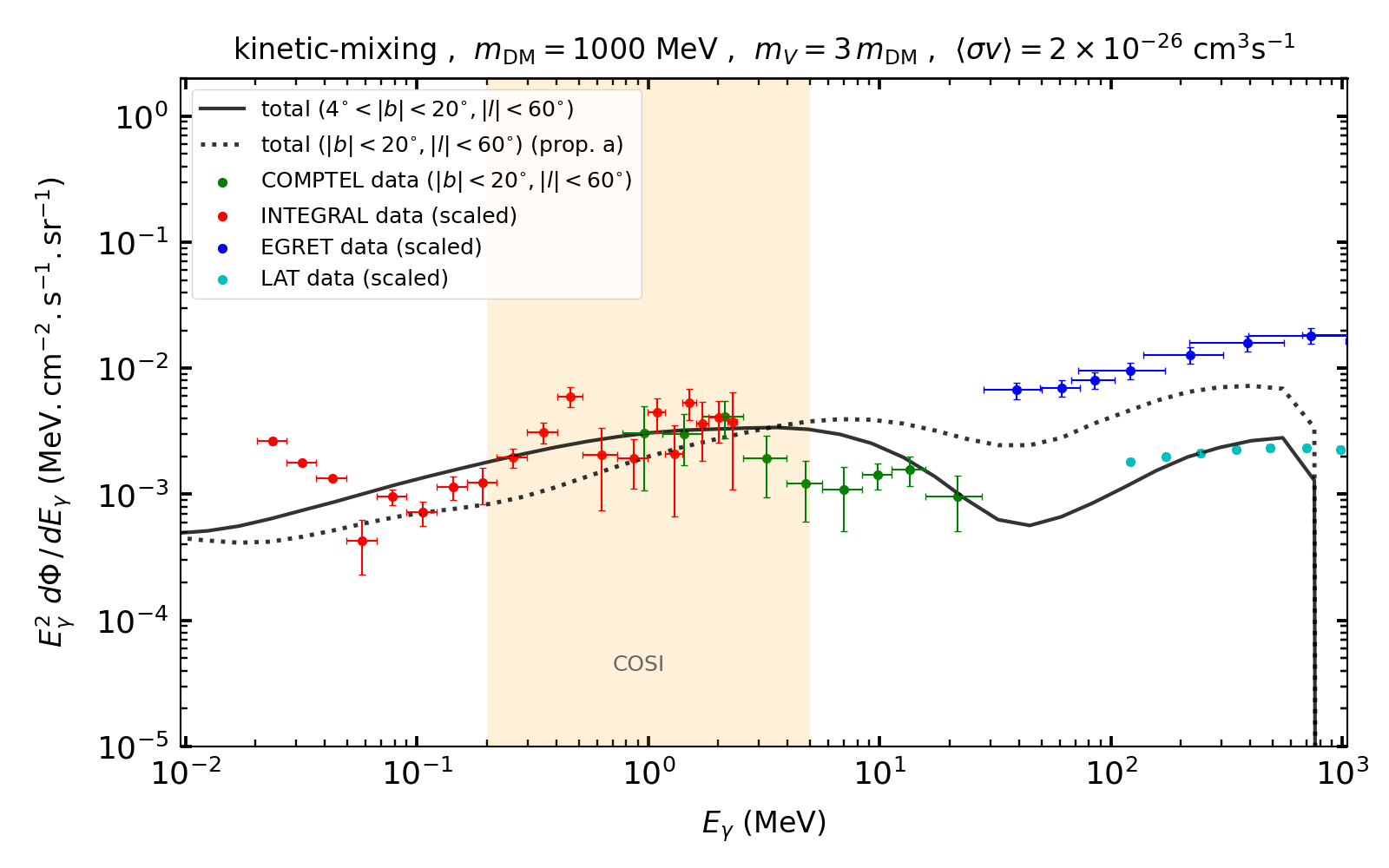}
\end{minipage}
\qquad
\begin{minipage}{0.27\textwidth}
\caption{\em Illustration of the {\bfseries effects of considering the full propagation} 
of the DM-induced $e^\pm$ in the Galaxy for three different DM masses: 10, 100 and 
1000 MeV. The model is the same one used in figure \ref{fig:DM_fluxes}. 
The galactic propagation of $e^\pm$ is solved numerically using 
the package \texttt{DRAGON2}, with the propagation model `prop. a' 
discussed in the text. For each DM mass, the total photon flux 
obtained (from a region $|b| < 20^{\circ} , |l| < 60^{\circ}$) 
in this way is denoted with a black dotted curve. 
This flux is compared to that obtained using the method used 
in figure \ref{fig:DM_fluxes} (the black solid curve), i.e.~using the semi-analytic approximated approach for the galactic propagation discussed in section \ref{sec:semi-analytic}, and masking the region $|b| \leq 4^{\circ}$.}
\label{fig:DM_fluxes_propagation}
\end{minipage}
\end{figure*}

In fig.~\ref{fig:DM_fluxes_propagation}, we show with black dotted 
curves the DM-induced total photon signals (prompt + secondaries) in the scenario 
where the $e^\pm$ distribution that gives rise to the secondaries are 
obtained considering the full galactic propagation (using the model `prop. a'). 
The three panels show the results for three different DM masses, 
$m_{\rm DM} = 10, 10^2$ and $10^3$ MeV, considering the kinetic-mixing model 
with $m_V = 3 \, m_{\rm DM}$ and 
$\langle \sigma v \rangle = 2\times10^{-26}$ $\rm cm^3 s^{-1}$. 
Such signals are estimated from the region $|b| < 20^{\circ} , |l| < 60^{\circ}$. 
For each mass, this total signal is compared to that 
(shown by the black solid line) obtained using the method discussed 
in sec.~\ref{sec:semi-analytic}, with a mask $|b| \le 4^{\circ}$. 
As we can see from this figure, in the considered photon energy range, 
the total photon signals (for different DM masses in the range MeV -- GeV) 
obtained in two different ways can differ at most by a factor of a few. 
In fact, in some cases, the total signal obtained considering the 
full propagation of $e^\pm$ remains comparatively stronger. 
In these cases the propagation phenomena, such as diffusion and re-acceleration, 
actually help in boosting the secondary signal flux 
coming from an extended region around the GC. 
A somewhat similar result was also obtained previously in ref.~\cite{DelaTorreLuque:2023olp}. 

Note that in fig.~\ref{fig:DM_fluxes_propagation} we show the effect of considering the 
galactic propagation of DM induced $e^\pm$ using the propagation model `prop.~a' as an illustration. 
We discuss the effects of considering both the above-mentioned propagation models 
in our main results presented in sec.~\ref{sec:results_and_discussion}.

\section{Cosmic-ray $e^\pm$ flux from sub-GeV DM annihilation in the Galaxy}
\label{sec:cosmic_ray_signal}

In this section we move to discuss the electron and positron signals from the annihilation of sub-GeV DM in the Galaxy.

\begin{figure*}[ht!]
\centering
\includegraphics[width=0.75\textwidth]{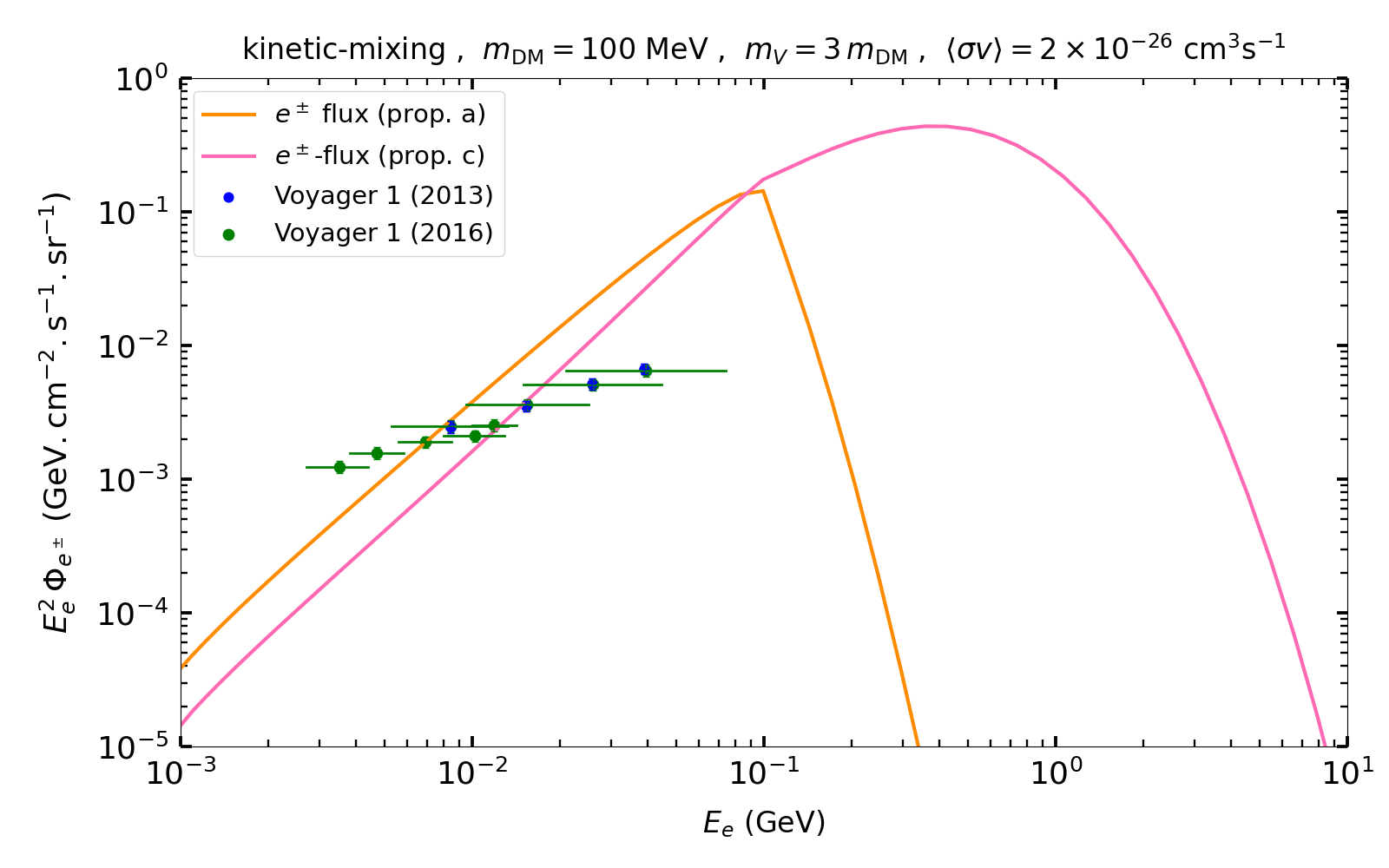}
\vspace{-3mm}
\caption{\em The DM-induced {\bfseries $\bm{e^\pm}$ flux} compared to the measurement of {\sc Voyager-1}. 
The two galactic propagation models `prop.~a' and `prop.~c' discussed in the text are considered. 
The DM model is the same vector portal one considered in fig.~\ref{fig:DM_fluxes}. 
The {\sc Voyager-1} 2013 data~\cite{Boudaud:2016mos, doi:10.1126/science.1236408} 
and {\sc Voyager-1} 2016 data~\cite{2016ApJ...831...18C} are shown by blue and 
green points, respectively.}
\label{fig:DM_cosmicray_flux}
\end{figure*}

The $e^\pm$'s propagate through the galactic environment 
and give rise to a contribution to the cosmic-ray flux near the solar system. 
We estimate such a flux of $e^\pm$ at the location of the {\sc Voyager-1} spacecraft 
and compare it with its data, for the ($e^{+} + e^{-}$) flux,  
to derive constraints on our considered DM models. 
Note that, since {\sc Voyager-1} is already 
outside the solar heliosphere, the effects of solar modulation 
(which suppresses the flux of low-energy $e^\pm$, as induced by sub-GeV DM) 
is negligible. 

The $e^\pm$ flux mentioned above is estimated 
(per energy, per area, per time and per solid angle) as: 
$\Phi_{e^\pm} = \frac{2 \, v_e}{4 \pi} \, \frac{dn_{e}}{dE_e}$, 
where $v_e$ and $\frac{dn_{e}}{dE_e}$ are, respectively, the velocity and 
the distribution of $e^\pm$ at the location of {\sc Voyager-1} 
(the factor 2 takes into account the contribution of $e^+$ and $e^-$). 
The distribution $\frac{dn_{e}}{dE_e}$ 
is obtained by solving the same propagation/transport equation mentioned 
in sec.~\ref{sec:full_propagation} (see eq. 2.1 of \cite{Evoli:2016xgn} and \cite{Evoli:2017vim}). 
We again obtain this using \texttt{DRAGON2}. 
For the DM density distribution we use the same profile of eq.~\ref{eq:rho_profile}. 

We use the model `prop.~a' as described in sec.~\ref{sec:full_propagation}. 
In addition, we also consider another model (named here {\bf `prop.~c'}) which 
refers to the model `DRE' from \cite{Orlando:2017mvd}. 
Note that `prop.~c' is almost the same as `prop.~a', 
the only difference being that the former includes a large 
Alfvén velocity, $v_A = 42.2$ km/s. 
As shown in \cite{Boudaud:2016mos, DelaTorreLuque:2023olp, Wang:2025jhy}, 
such a modification in $v_A$ should have a significant effect 
on the $e^\pm$ flux at the {\sc Voyager}'s location. 

\medskip

In fig.~\ref{fig:DM_cosmicray_flux} we illustrate the resulting DM-induced 
$e^\pm$ fluxes. These fluxes are shown considering the kinetic-mixing model 
with $m_{\rm DM} = 100$ MeV, $m_V = 3 \, m_{\rm DM}$ and 
$\langle \sigma v \rangle = 2\times10^{-26}$ $\rm cm^3 s^{-1}$. 
The {\sc Voyager-1} data \cite{Boudaud:2016mos, doi:10.1126/science.1236408, 2016ApJ...831...18C} 
are also shown for comparison. 
As it can be seen, due to the presence of a 
higher re-acceleration (corresponding to a larger value of $v_A$), 
the model `prop.~c' gives rise to an $e^\pm$ flux that 
is comparatively more energetic.

\section{Data and analysis} 
\label{sec:obs_data_analysis}

We obtain constraints on the sub-GeV DM models by comparing 
the DM-induced total photon and the cosmic-ray signals (discussed in the previous sections) 
with the existing $X$-ray/$\gamma$-ray data from {\sc Integral}, {\sc Comptel} and {\sc Fermi-Lat}, and the cosmic-ray $e^\pm$ data from {\sc Voyager-1}, respectively. 
These observations are briefly presented below~\footnote{We also considered the $\gamma$-ray 
observations by {\sc Egret}, but, since this observation is expected to 
provide a comparatively weaker constraint on the DM annihilation signal 
(see, for example, fig. \ref{fig:DM_fluxes} or fig. \ref{fig:DM_fluxes_propagation}), 
we do not discuss it here and henceforth.}. 

\subsection{Observed data}
\label{sec:data}

\begin{itemize}
\item[$\circ$] \underline{{\sc Integral}}: 
We use the data from the observations of $\sim20$ keV -- 2.4 MeV photons from 
the region $|b| < 15^{\circ} , |l| < 30^{\circ}$ around the GC. 
These data, obtained with the {\sc Spi} instrument onboard {\sc Integral}, are 
taken from ref.~\cite{2011ApJ...739...29B}. For this observation we assume an 
energy resolution of $0.2\%$ (which is the typical energy resolution of this telescope). 
We also consider as our fiducial background model 
the possible diffuse photon flux (discussed in \cite{2011ApJ...739...29B}) 
that arises due to different astrophysical phenomena and 
can explain this photon observation. 
\item[$\circ$] \underline{\sc Comptel}: 
We consider the observation of {\sc Comptel} from a region $|b| < 20^{\circ} , |l| < 60^{\circ}$. 
The data, in the photon energy range 
$\sim1-15$ MeV, are taken from \cite{Essig:2013goa,KappadathCOMPTEL}. 
The energy resolution considered for this telescope is $5\%$ (see, e.g., \cite{Coogan:2019qpu}). 
We also consider the diffuse astrophysical background photon flux from ref.~\cite{Bartels:2017dpb}. 
This is modeled as a power-law including a super-exponential cutoff, and is in reasonable agreement 
with the {\sc Comptel} data from a region $|b| < 5^{\circ} , |l| < 30^{\circ}$. 
We scale this diffuse background to the ROI $|b| < 20^{\circ} , |l| < 60^{\circ}$ and 
use as our fiducial background flux. Note that, while the {\sc Comptel} data have not been included in past works such as ref.~\cite{Cirelli:2020bpc,Cirelli:2023tnx}, they actually turn out to be pertinent, albeit in a narrow window of the parameter space. 
\item[$\circ$] \underline{\sc Fermi-Lat}: 
We also consider the {\sc Fermi-Lat} observation from a region of radius $10^\circ$ around 
the GC. For our purposes, we use the data in the energy range 
$100 \, {\rm MeV} \lesssim E_\gamma \lesssim 1$ GeV provided in~\cite{Orlando:2017mvd, 2017ApJ...840...43A}. 
We assume an 
energy resolution of $7.5\%$ \cite{Coogan:2019qpu}. 

\item[$\circ$] \underline{\sc Voyager1}: 
We consider the data for the $e^+ + e^-$ flux collected 
by the {\sc Voyager-1} spacecraft outside the solar heliosphere. 
These data, corresponding to {\sc Voyager-1} (2013) over an $e^\pm$ energy range $\sim8-40$ MeV 
and to {\sc Voyager-1} (2016) over the range $\sim3-100$ MeV,  
are taken from \cite{Boudaud:2016mos, doi:10.1126/science.1236408} 
and \cite{2016ApJ...831...18C}, respectively. 
For each DM model, at each DM mass, we consider the strongest bound 
obtained from either of the two data sets as our {\sc Voyager-1} limit. 
\end{itemize} 

\subsection{Analysis with the observed data}
\label{sec:analysis_obs_data}

We compare different DM signals with the corresponding experimental observations 
discussed above by taking the following two analysis approaches. 

\begin{itemize}
\item[$\blacktriangleright$] \underline{\bf Conservative approach}: 
In this approach we follow a strategy similar to the one used in ref.~\cite{Cirelli:2009vg, Essig:2013goa, Boudaud:2016mos}. 
We do not assume anything on the astrophysical backgrounds and compare the DM-induced signals directly to the observed data. 
For each experimental observation with the corresponding data points, we impose the constraint 
that the annihilation signal from a fixed DM model (for a given DM mass) should not 
exceed any of the experimental data points at any energy bin by more than $2\sigma$. 
For example, for each $X$-ray/$\gamma$-ray observation from a particular observation region, 
we make sure that, for a given DM mass, the DM-induced total photon signal (prompt + secondaries) 
(discussed in sec.~\ref{sec:photon_signal}) from that region does not exceed 
the observed central value plus twice the error bar in any energy bin. 
Also, in the context of the cosmic-ray $e^\pm$ data observed by {\sc Voyager-1}, 
we take the same approach to constrain the signal from DM.  
These lead to upper-limits on the total annihilation cross-section of DM in a given model, 
which are presented in sec.~\ref{sec:results_and_discussion}. 

\item[$\blacktriangleleft$] \underline{\bf Optimistic approach}: 
In this alternative approach we assume the observed data to be of some astrophysical origin, and then impose the condition that the DM signal, when added to such astrophysical background, 
does not exceed the data points. This makes our constraints on DM comparatively stronger. We do this for {\sc Integral} and {\sc Comptel} data only, for which we possess a reliable estimate of the background (see section~\ref{sec:data}).~\footnote{In the case of the {\sc Fermi-Lat} GC observation, the modeling of the corresponding photon background is quite uncertain (see e.g. \cite{Orlando:2017mvd}) and subject of intense investigations in itself so that entering in the details would go beyond our scope. 
The {\sc Voyager-1} observations also exhibit a relatively large uncertainty in 
the corresponding background modeling (see e.g.~\cite{Boudaud:2016mos}).} 
We adopt a simple chi-squared statistic: 
\begin{equation}
\chi^2 = \sum_{i} \, 
\frac{(\phi^0_i - \phi^{\rm DM}_i(\langle \sigma v \rangle) - \phi^{\rm BG}_i)^2}{\sigma^2_i} \, ,
\end{equation}
where $\phi^0_i$ is the observed data from a given experiment (at the $i$-th energy bin) with 
$\sigma_i$ representing the corresponding error bar, $\phi^{\rm BG}_i$ is the 
astrophysical background that fits the corresponding observed data, and 
$\phi^{\rm DM}_i(\langle \sigma v \rangle)$ is the signal from the annihilation of 
DM (with a fixed mass) in a given model.  
For a given experiment, we obtain a 95\% C.L. upper-limit on the 
signal normalization $\langle \sigma v \rangle$ (for a given $m_{\rm DM}$) 
by requiring that $\Delta\chi^2 = \chi^2 - \chi^2_0 = 2.71$, where $\chi^2_0$ is the 
best-fit $\chi^2$-value with zero DM signal. 
In sec.~\ref{sec:results_and_discussion} we present 
such constraints. 
\end{itemize}

\subsection{Upcoming MeV telescope {\sc Cosi}} 
\label{sec:MeV_telescope}

As we saw in sec.~\ref{sec:photon_signal}, the photon signals 
arising from the annihilation of DM particles 
with masses in the range MeV - GeV 
populate the energy range $\sim0.1-100$ MeV, 
the so called `MeV gap'~\cite{Engel:2022bgx}. 
Such a range will be probed by a number of upcoming space-based MeV telescopes, 
e.g., {\sc Cosi}~\cite{Tomsick:2019wvo, Beechert:2022phz}, 
{\sc Amego}~\cite{AMEGO:2019gny, 2020SPIE11444E..31K}, 
{\sc e-Astrogam}~\cite{e-ASTROGAM:2016bph, e-ASTROGAM:2017pxr}, 
{\sc Gecco}~\cite{Orlando:2021get} 
{\sc Grams}~\cite{Aramaki:2021o5}, 
{\sc Adept}~\cite{Hunter:2013wla} and 
{\sc Pangu}~\cite{Wu:2014tya} with sensitivities 
much better than that of the existing or past telescopes 
(such as {\sc Integral}, {\sc Comptel} and {\sc Egret}).  

In this work, we focus on 
the upcoming telescope {\sc Cosi}, which has already been selected to fly. 
{\sc Cosi} is a space-based wide field-of-view telescope 
with an operational energy range $0.2 - 5$ MeV. 
It employs a compact Compton telescope design and 
contains several high-purity germanium semiconductor detectors. 
These help {\sc Cosi} have a high-resolution spectroscopy 
(with an energy resolution $\lesssim 4\%$), 
a direct imaging over a wide field-of-view ($\gtrsim 25\%$ of the sky), 
as well as an effective suppression for the background events. 
Its effective area and angular resolution are above a few ${\rm cm^2}$ 
and above a few degrees, respectively. 
With its scheduled launch in 2027, {\sc Cosi} is supposed to start surveying the sky 
orbiting Earth and will be able to assure a daily full-sky coverage. 
For an overview of its instrumentation and technical details, 
see~\cite{Tomsick:2019wvo, Beechert:2022phz}. 

We estimate the projected sensitivity of {\sc Cosi} on our considered sub-GeV 
DM models following a methodology similar to the one 
adopted in \cite{Caputo:2022dkz} 
(regarding the continuum photon signal from the GC). 
Note that ref.~\cite{Caputo:2022dkz} studied the prospects of {\sc Cosi} 
for sub-GeV DM annihilation (considering a couple of annihilation channels) 
based on only the prompt emission. 
We consider the same target region (as in \cite{Caputo:2022dkz}) 
of radius $10^\circ$ around the GC and compare the 
DM-induced total photon signal composed of prompt and secondary emissions 
from this region with the {\sc Cosi} sensitivity, in order to estimate 
the projection on the corresponding DM annihilation rate. 
The $3\sigma$ projected sensitivity of {\sc Cosi} corresponding 
to the continuum emission is obtained from figure 4 of ref.~\cite{Negro:2021urm}. 
This {\sc Cosi} sensitivity (for a 2 yrs of mission time), 
provided for a disk-like observation 
region with $20^\circ$ radius around the GC, is then rescaled to 
our observation region 
following the prescription given in \cite{Negro:2021urm} (as also used in \cite{Caputo:2022dkz}). 
Knowing that such sensitivity scales as the squared root of the observation time, 
we obtain the {\sc Cosi} sensitivity for a mission time 
of 4 yrs (which corresponds to approximately 1 yr of observation time, 
see \cite{Negro:2021urm}). Note that the minimal, nominal mission lifetime of {\sc Cosi} is 2 years, but a longer duration is likely.
We then obtain the projected sensitivity on the total DM annihilation rate 
for a given DM mass and a given model by ensuring that 
the corresponding total DM signal overcomes the $3\sigma$ {\sc Cosi} sensitivity 
at least in one of the five energy bins for which the sensitivity 
is provided in ref.~\cite{Negro:2021urm}. The projections obtained in this way are presented 
in sec.~\ref{sec:results_and_discussion}.

\section{Results and discussion}
\label{sec:results_and_discussion}

\subsection{Constraints from existing data} 
In this section we present the indirect detection constraints 
obtained on the two considered sub-GeV DM models based on existing astrophysical data 
and the two approaches discussed in sec.~\ref{sec:obs_data_analysis}. 
These constraints are shown on the total annihilation cross-section 
$\langle \sigma v \rangle$ as a function of $m_{\rm DM}$ in the range 1 MeV -- 1 GeV. 

\begin{figure*}[!t]
\hspace{-1.8cm}
\begin{minipage}{0.61\textwidth}
\includegraphics[width=\textwidth]{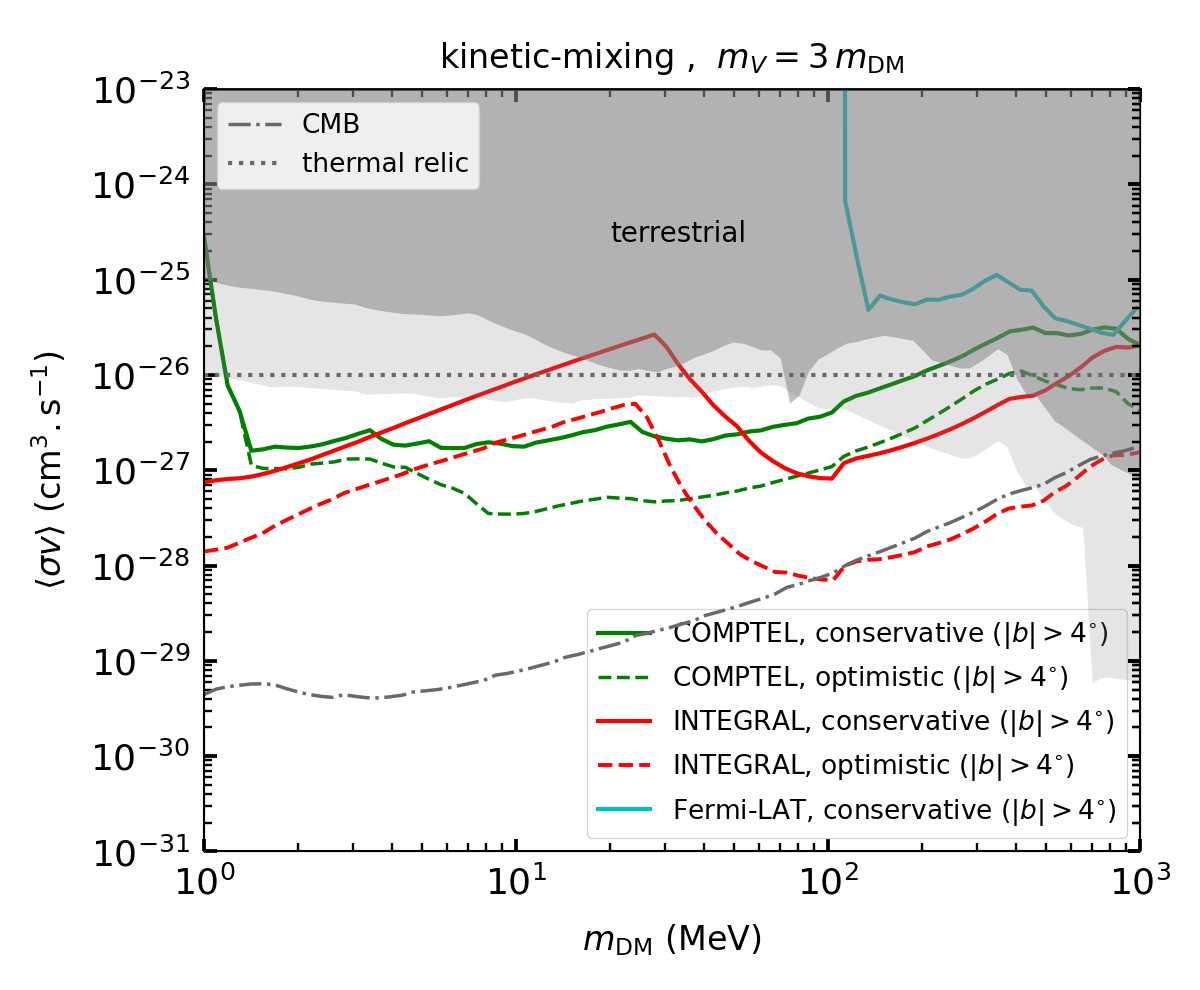}\\
\includegraphics[width=\textwidth]{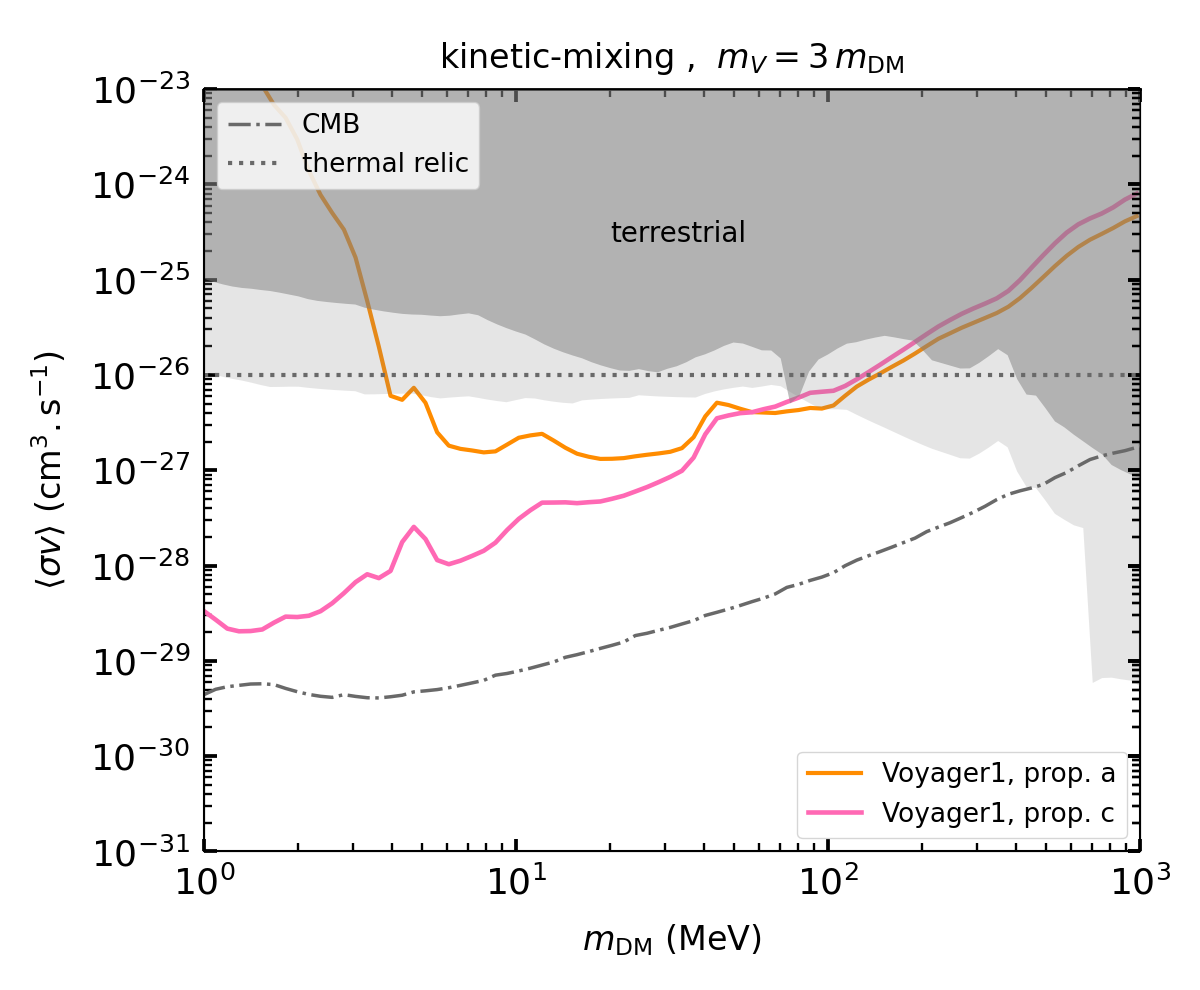}
\end{minipage}
\begin{minipage}{0.61\textwidth}
\includegraphics[width=\textwidth]{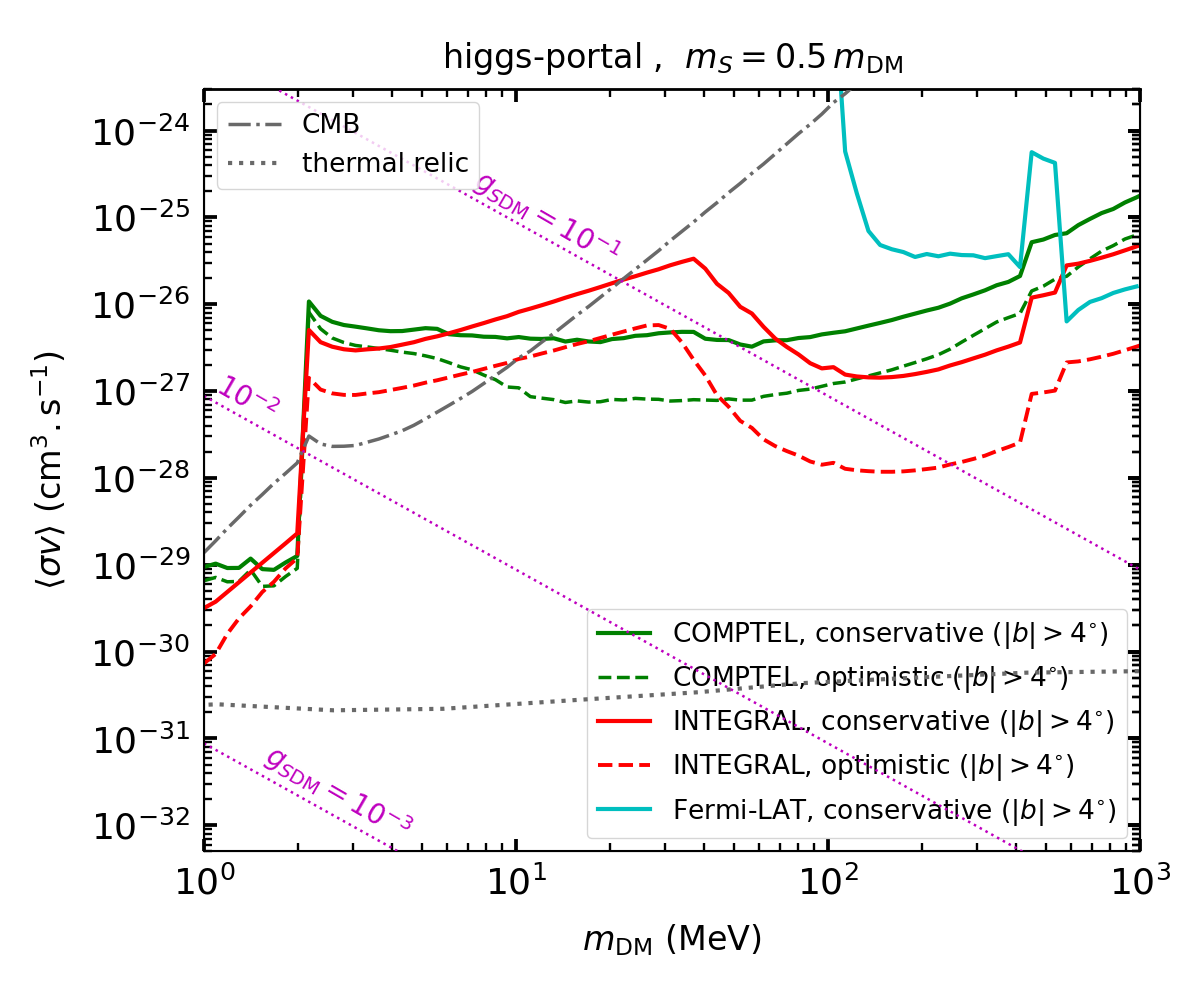}\\
\includegraphics[width=\textwidth]{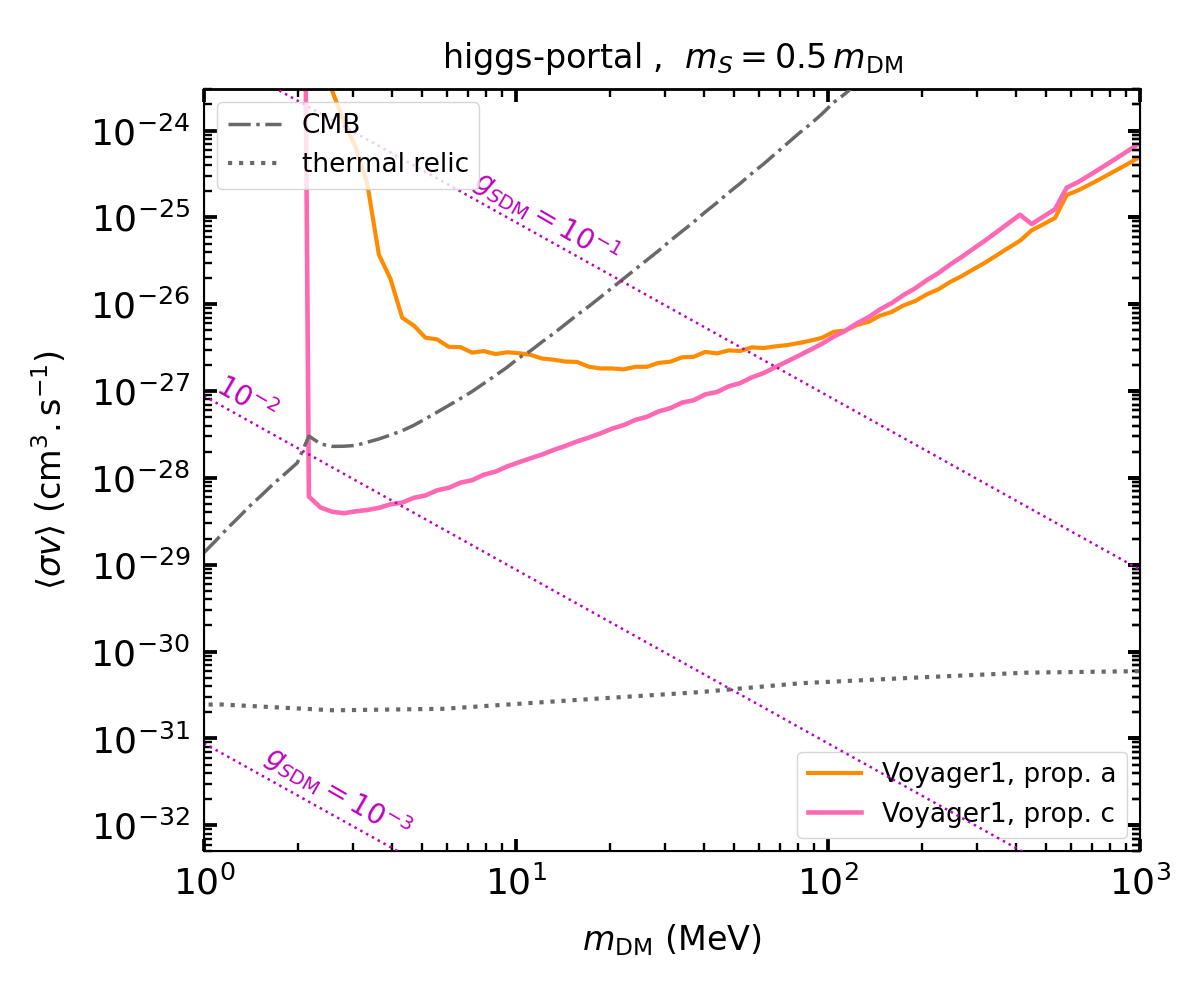}
\end{minipage}
\caption{\em {\bfseries Indirect detection upper-limits} obtained on the total $\langle \sigma v \rangle$ 
of the two considered DM models (vector portal and scalar portal, left and right panels respectively) based on existing astrophysical 
observations. Limits obtained by comparing the DM-induced total photon 
signal (primary + secondaries) with data from {\sc Comptel}, 
{\sc Integral} and {\sc Fermi-Lat} are presented in the upper panels by green, red and cyan curves, respectively. 
The solid curves are obtained under the conservative analysis, 
while the dashed curves are obtained using the optimistic analysis. 
Limits obtained by comparing the DM-induced 
$e^\pm$ flux with {\sc Voyager-1} data (under the conservative analysis) 
are shown in the lower panels by the orange and pink solid curves, 
which correspond to the galactic propagation models 
`prop.~a' and `prop.~c', respectively 
(see the text). We also show the thermal relic DM line (gray dotted lines) and the CMB bounds (gray dashed-dotted lines), as well as the combined exclusions (dark gray region) and 
the combined projections (light gray region) from terrestrial experiments \cite{Krnjaic:2022ozp}. 
The contours corresponding 
to different values of the 
coupling $g_{S{\rm DM}}$ are 
indicated by the purple dotted lines on the right. 
\label{fig:sv_mx_lim}}
\end{figure*}

\medskip

Fig.~\ref{fig:sv_mx_lim} shows the results for the kinetic-mixing and 
higgs-portal models in the left and right panels, respectively. 
In the top row of the figure, the red, green and cyan 
curves show the constraints obtained on the two models 
from $X$-ray/$\gamma$-ray observations of {\sc Integral}, {\sc Comptel} and {\sc Fermi-Lat}, 
respectively. The total photon signal is obtained using the method described 
in sec.~\ref{sec:semi-analytic}, with a mask $|b| \leq 4^{\circ}$. 
In each case, the constraint shown by the solid curve 
is obtained under the conservative analysis 
described in sec.~\ref{sec:analysis_obs_data}, i.e., comparing the DM signal 
directly with the observed data without making any assumption 
on the astrophysical background. The strong {\sc Integral} limit between $\sim 30$ MeV and 1 GeV is a consequence of a strong secondary signal, in particular the one produced via the ICS of the DM-induced $e^\pm$ off the optical starlight,  at an energy where the intensity of the {\sc Integral} data is at a minimum (see figs.~\ref{fig:DM_fluxes} and \ref{fig:DM_fluxes_propagation} 
for an illustration). 
For $m_{\rm DM} \gtrsim 50$ MeV, {\sc Integral} typically provides 
the dominating constraint on sub-GeV DM among different photon observations 
based on mainly such ICS signals. For the higgs-portal case, the {\sc Fermi-Lat} 
constraint becomes marginally stronger for $m_{\rm DM} > 600$ MeV. 
For $m_{\rm DM}$ below $\sim30$ MeV, the ICS signal falls below 
the {\sc Integral} energy range, and thus the corresponding {\sc Integral} constraint 
is driven by the signals resulting from bremsstrahlung, prompt emission and 
in-flight annihilation. {\sc Comptel} provides a better constraint in the 
mass range from a few MeV to $\sim60$ MeV, based on mainly bremsstrahlung, 
prompt emission and in-flight annihilation. 
Below this mass range {\sc Integral} dominates again. 
The plunge in the {\sc Integral/Comptel} limit in the 
higgs-portal case for $m_{\rm DM}$ below the electron mass threshold (i.e., $m_{\rm DM}<2$ MeV) 
results due to the photon line signals appearing from the 
decay of the mediator $S$ into prompt $\gamma$'s 
(which is the dominating decay mode of $S$ for these DM masses). 

\medskip

In fig.~\ref{fig:sv_mx_lim} the constraints shown by the colored 
{\em dashed} curves (for {\sc Integral} and {\sc Comptel}) are obtained 
using the optimistic analysis taking into account 
the standard photon backgrounds mentioned in sec.~\ref{sec:obs_data_analysis}. 
Adopting this approach, the 
bounds are improved significantly, by an amount that depends on the $m_{\rm DM}$ value. 
For example, in the case of {\sc Integral}, 
this improvement is an order of magnitude for $m_{\rm DM} \gtrsim 80$ MeV. 
Due to this, it is possible to constrain 
the parameter space of the kinetic-mixing/higgs-portal model down to 
$\langle \sigma v \rangle \simeq 10^{-28} \, {\rm cm}^3/{\rm s}$ 
for a DM mass of $\sim100$ MeV. 

\medskip

In the lower row of fig.~\ref{fig:sv_mx_lim} we present 
the exclusion limits obtained on the kinetic-mixing and higgs-portal models 
from {\sc Voyager-1} data. The orange and pink curves show the limits 
obtained considering the two galactic propagation 
models `prop.~a' and `prop.~c' (mentioned in secs.~\ref{sec:photon_signal} 
and \ref{sec:cosmic_ray_signal}), respectively. For DM masses below $\sim50$ MeV, 
model `prop.~c' leads to a much stronger constraint compared to 
that obtained with model `prop.~a'. As noted in sec.~\ref{sec:cosmic_ray_signal}, 
since `prop.~c' includes a larger re-acceleration velocity ($v_A$), 
for a given $m_{\rm DM}$ this model gives rise to a much more energetic 
cosmic-ray flux of $e^\pm$ compared to that obtained with model `prop.~a'. 
As a result, the soft $e^\pm$ spectra induced by a low mass DM get boosted 
in energy after the propagation and fall in the energy range of {\sc Voyager-1}, 
leading to a stronger constraint on that DM mass. 
Based on such effects, using {\sc Voyager-1} data one can exclude 
$\langle \sigma v \rangle$ for both models down to 
$\sim10^{-28} \, \rm cm^3/s$ for a DM mass of $\sim10$ MeV. 
For the higgs-portal model, since $m_S = m_{\rm DM}/2$ in our setup, the {\sc Voyager-1} limits 
evaporate for $m_{\rm DM}<2$ MeV due to the closing of 
all leptonic annihilation channel including $e^+e^-$.  

\medskip

In fig.~\ref{fig:sv_mx_lim_future}, we show the combined upper-limits 
from different existing $X$-ray/$\gamma$-ray observations 
with the black solid curves (corresponding to the conservative approach), 
together with the bounds from {\sc Voyager-1} in pink curves 
(corresponding to the propagation model `prop.~c'). 
As we can see, for both DM models, the existing photon observations provide 
better constraints compared to those coming from {\sc Voyager-1} 
for $m_{\rm DM} \gtrsim$ a few tens of MeV. Below this mass range, {\sc Voyager-1} 
usually provides a better constraint (except for the higgs-portal case with $m_{\rm DM}\lesssim2$ MeV) 
depending on the choice of the propagation model.

\subsection{Projected sensitivity of {\sc Cosi}} 
\label{sec:projectedCOSI}

\begin{figure*}[!t]
\hspace{-1.8cm}
\begin{minipage}{0.61\textwidth}
\includegraphics[width=\textwidth]{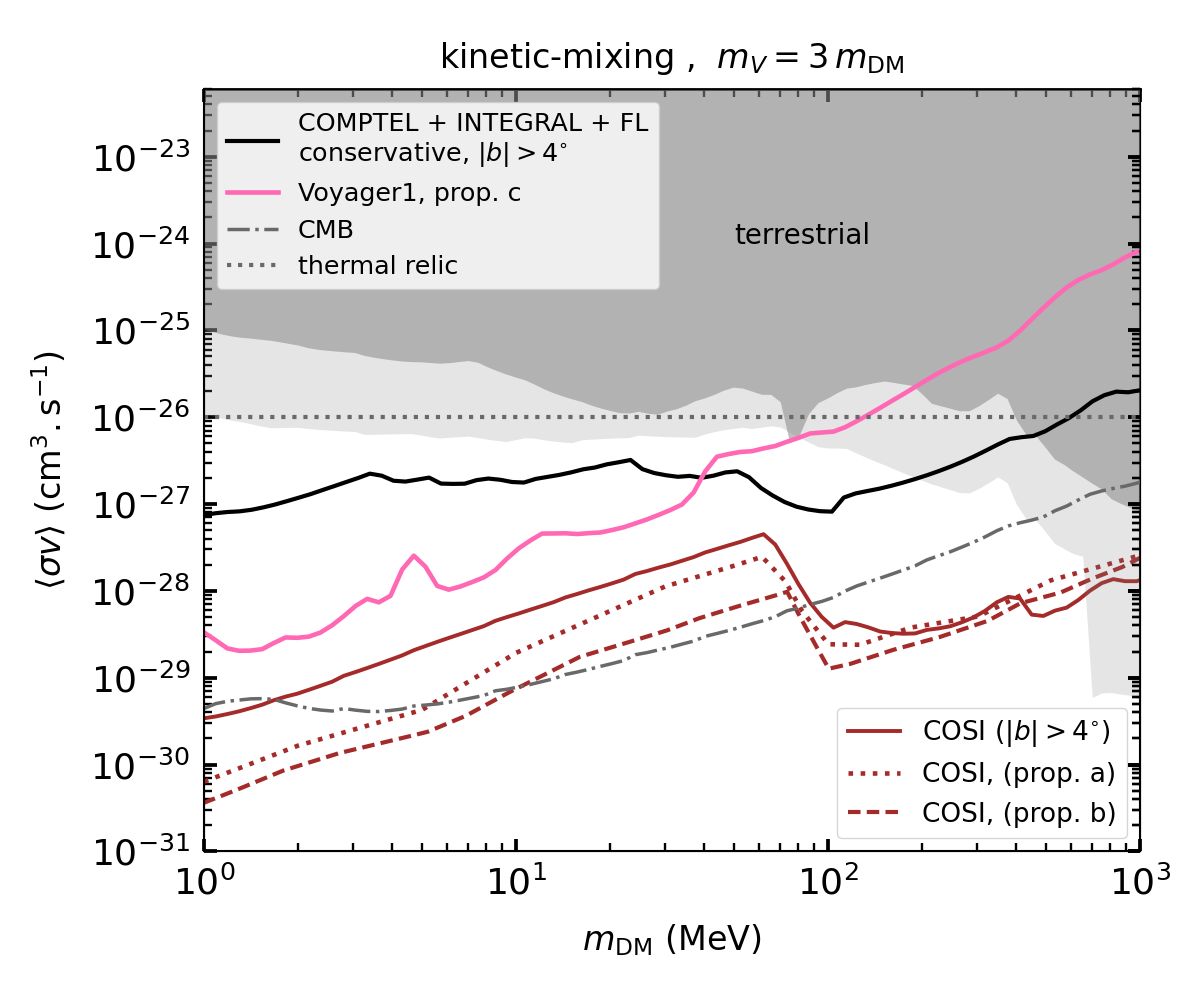}
\end{minipage}
\begin{minipage}{0.61\textwidth}
\includegraphics[width=\textwidth]{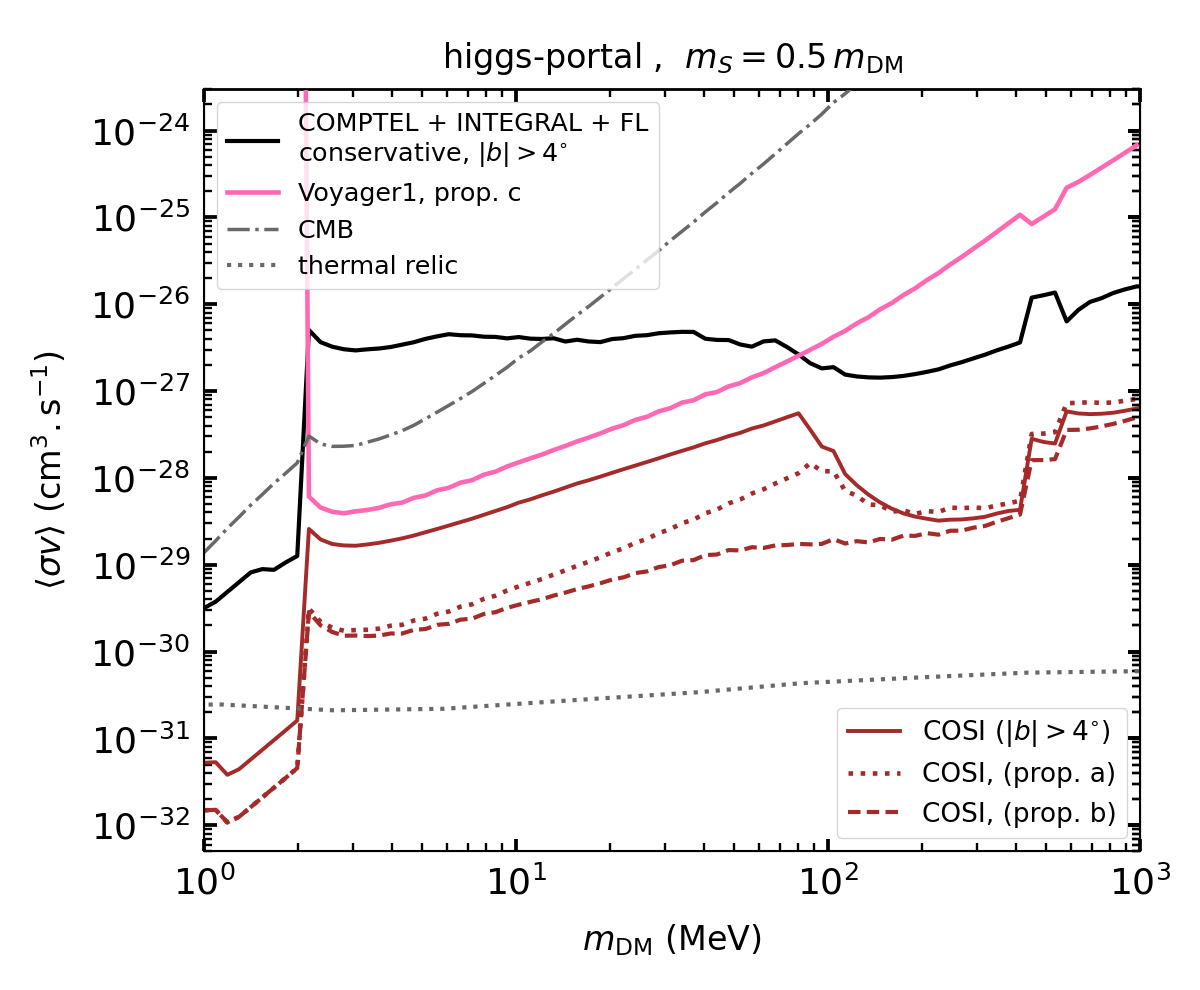}
\end{minipage}
\caption{\em {\bfseries Summary of the constraints} on the kinetic-mixing model (left panel) 
and the higgs-portal model (right panel) as well as $3\sigma$ {\bfseries projected sensitivities} of the upcoming MeV telescope {\sc Cosi} (for an observation time of 1 year). The latter ones are shown by the brown curves. The brown solid curves are the projections where 
total photon signal is obtained using the method described 
in sec.~\ref{sec:semi-analytic} (with a mask $|b| \leq 4^{\circ}$). 
The effects of considering the full Galactic propagation of $e^\pm$ are shown 
by the brown dotted (dashed) curves which correspond to 
the propagation model `prop.~a' (`prop.~b') discussed in sec.~\ref{sec:full_propagation}. 
} 
\label{fig:sv_mx_lim_future}
\end{figure*}
In this section we discuss the projected sensitivities of the upcoming 
MeV telescope {\sc Cosi} in probing the sub-GeV DM models based on the 
approach described in sec.~\ref{sec:MeV_telescope}. 
In fig.~\ref{fig:sv_mx_lim_future} we show with brown curves 
these {\sc Cosi} sensitivities (at $3\sigma$) in the usual 
$\langle \sigma v \rangle - m_{\rm DM}$ plane for kinetic-mixing (left panel) 
and higgs-portal (right panel) models. 
As mentioned in sec.~\ref{sec:MeV_telescope}, such sensitivities are obtained 
considering 1 year of {\sc Cosi} observation towards a $10^\circ$ disk region around the GC. 

The brown solid curves in fig.~\ref{fig:sv_mx_lim_future} correspond to 
the projections obtained using the total photon signals estimated following 
sec.~\ref{sec:semi-analytic} with a mask $|b| \leq 4^{\circ}$. 
The brown dotted and dashed curves, on the other hand, show the projections 
obtained with the photon signals estimated considering the effect of 
galactic propagation of DM-induced $e^\pm$, under the models 
`prop.~a' and `prop.~b', respectively (discussed in sec.~\ref{sec:full_propagation}). 
As pointed out in sections~\ref{sec:semi-analytic} and \ref{sec:full_propagation}, 
the cases with propagation can actually produce a larger flux for the total photon signal 
over the energy range of interest, especially for $m_{\rm DM}$ below a 
few hundreds of MeV. As a result, in this mass range, the {\sc Cosi} projections 
are comparatively stronger. Model `prop.~b', in spite of including the convection of $e^{\pm}$, 
provides a stronger projection (especially for $m_{\rm DM}$ 
between a few MeV and $\sim200$ MeV) due to its lower value for 
the diffusion coefficient and a larger value for 
the re-acceleration velocity (see sec.~\ref{sec:full_propagation}). 
We recall from sec.~\ref{sec:MeV_telescope} that these propagation models 
were used in ref.~\cite{Orlando:2017mvd} as baseline models for studying 
the secondary photon emissions from astrophysical processes towards the GC. 
Such galactic emissions were then used in ref.~\cite{Negro:2021urm} 
as fiducial backgrounds while estimating the {\sc Cosi} sensitivity 
(which is considered in the present work). Hence this ensures consistency between out treatment of the signal and of the background, for these cases.

Like in the case of the $X$-ray/$\gamma$-ray upper-limits discussed above, 
the strengthening in the {\sc Cosi} projections around $\sim100$ MeV and above is due to 
the excess of the ICS signal produced by the scattering of 
DM-induced $e^\pm$ off the optical starlight, at an energy 
where the sensitivity of {\sc Cosi} is stronger~\cite{Negro:2021urm}. 
I.e., for $m_{\rm DM} \gtrsim 100$ MeV the {\sc Cosi} sensitivity is driven 
mainly by the ICS signal, while for $m_{\rm DM} < 100$ MeV the main photon 
signal is a combination of bremsstrahlung, in-flight annihilation and prompt emission. 
For the higgs-portal model, the plunge in the projections 
for $m_{\rm DM}$ below the electron threshold ($m_{\rm DM} < 2$ MeV) 
appears due to the photon line signals produced from the decay of the mediator.
Also, note that the visible features at higher DM masses for both the models 
correspond to the muon and pion thresholds. 

\bigskip

As can be seen from fig.~\ref{fig:sv_mx_lim_future}, for both the 
sub-GeV DM models, the upcoming {\sc Cosi} observations will be able to probe 
a vast region of the parameter space that lies beyond the constraints 
placed by the existing $X$-ray, $\gamma$-ray and cosmic-ray data. 
For example, for the kinetic-mixing model, {\sc Cosi} is sensitive to 
such values of $\langle \sigma v \rangle$ that are at 
least an order of magnitude below the combined bound set by existing 
indirect searches for $m_{\rm DM} \lesssim 40$ MeV and $m_{\rm DM} \gtrsim 80$ MeV. 
For the higgs-portal model, such improvements vary from a factor of few to 
one/two orders of magnitude, depending on the choice of propagation 
parameters used for the estimates related to {\sc Cosi}.  

\subsection{Importance of secondary photons} 
\begin{figure*}[!t]
\hspace{-2.5mm}
\includegraphics[width=0.51\textwidth]{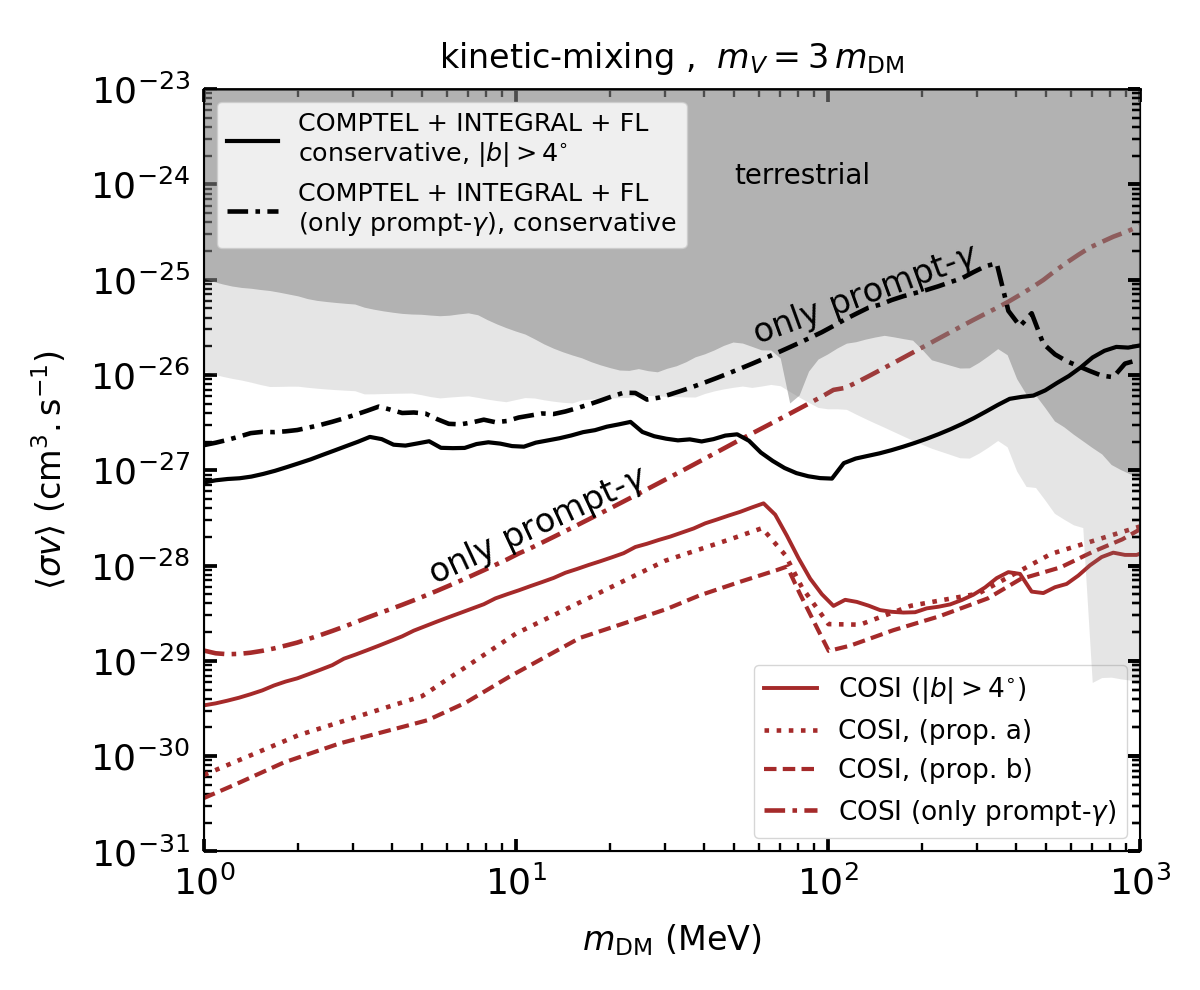}
\hspace{-2mm}
\includegraphics[width=0.51\textwidth]
{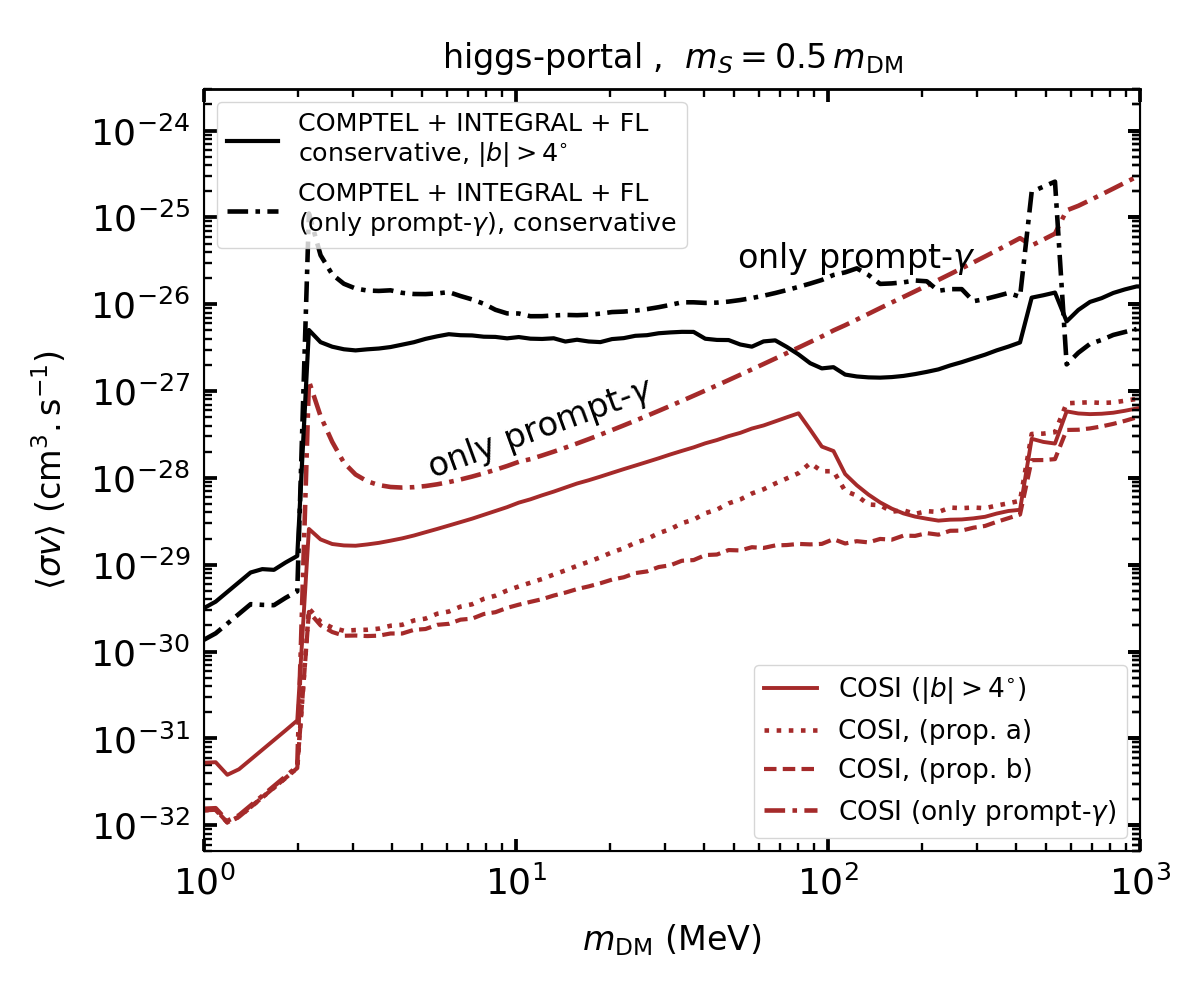}\\
\vspace{-6mm}
\caption{\em {\bfseries Importance of the inclusion of secondary photons}. 
The combined conservative bounds and the {\sc Cosi} projections are shown with 
and without including the secondaries, considering the 
kinetic-mixing (left panel) and the higgs-portal (right panel) models. 
The dashed-dotted black and brown curves 
correspond to the case neglecting the secondary 
photon signals, while the other black and brown curves show the results obtained 
including the secondaries and are same as the ones presented in fig.~\ref{fig:sv_mx_lim_future}. 
}
\label{fig:sv_mx_lim_secondaries}
\end{figure*}

An important ingredient of the present work
consists in constraining or probing sub-GeV DM based on signals that includes different secondary emissions produced by DM-induced $e^\pm$ in the Galaxy. 
To stress this point, in fig.~\ref{fig:sv_mx_lim_secondaries} we illustrate the 
importance of including such secondary emissions for the two considered sub-GeV DM models. 
For each model, the solid black line in fig.~\ref{fig:sv_mx_lim_secondaries} 
corresponds to the combined bound (obtained with the conservative approach) 
from existing $X$-ray/$\gamma$-ray observations 
and is the same as the one shown in fig.~\ref{fig:sv_mx_lim_future}. 
The black dashed-dotted line, on the other hand, corresponds to the bound 
obtained by switching off all the secondary emissions 
and considering only the prompt $\gamma$-ray emission (without the mask $|b| \le 4^\circ$). 
As it can be seen, the bounds including the secondaries are stronger 
for almost all the DM mass range, although the effect is much more prominent 
in the mass range $\sim50 - 600$ MeV where the improvement can be of one order of magnitude. 
Indeed, in this mass range, the total signal is driven mainly by the ICS signal. 
For completeness, we also notice that, for some DM masses e.g.~close to 1 GeV, the limits obtained considering both prompt and secondary photon signals turn out to be weaker 
compared to those obtained using only prompt $\gamma$-ray signal. 
This is mainly due to the fact that, in the former case  
as mentioned in sec.~\ref{sec:semi-analytic}, 
we remove all the signals (both prompt and secondaries) coming from a $4^\circ$ region 
above and below the GP (to be on the conservative side), 
while in later case the prompt signal from this region is included in the analysis. 

The effect of considering the secondaries is much more significant for the {\sc Cosi} projections, 
as can be seen by comparing the brown dashed-dotted lines which do not include secondaries 
with the other brown lines which include secondaries and are the same as the ones 
shown in fig.~\ref{fig:sv_mx_lim_future}. Again the significance of secondaries 
turns out to be more important for higher DM masses (in particular for $m_{\rm DM} \gtrsim 100$ MeV) 
where the improvement can be significantly large, up to a few orders of magnitude. 
For example, for the kinetic-mixing model, at $m_{\rm DM} \simeq 500$ MeV, 
the projection obtained including secondaries (with a mask $|b| \le 4^\circ$) 
is almost three orders of magnitude stronger than that obtained without secondaries 
(without the mask $|b| \le 4^\circ$). As mentioned above, such an improvement 
relies mainly on the inclusion of the ICS signal. 
Note that, for some DM masses, the projections based on the total photon signal 
can be weaker compared to that obtained with only prompt emission 
(see, e.g., the higgs-portal case for $m_{\rm DM} < 2$ MeV) due to the same reason 
mentioned above in the case of the existing photon observations.  

\subsection{Comparison with other constraints}

\subsubsection{Thermal freeze-out relic DM}
Although this work does not require specifically a thermal freeze-out 
relic DM scenario, we show, just for the sake of comparison, 
in each plot of the $\langle \sigma v \rangle - m_{\rm DM}$ plane 
the contour (by the gray dotted line) that yields the correct 
DM relic density~\cite{Planck:2018vyg} 
under the standard thermal freeze-out scenario. 
For the kinetic-mixing model, where the DM annihilation proceeds 
through $s$-wave, this $\langle \sigma v \rangle$ 
for the thermal relic DM is taken to be 
$\langle \sigma v \rangle \simeq 10^{-26}$ $\rm cm^3 s^{-1}$ 
\cite{Coogan:2021rez, Coogan:2021sjs, Steigman:2012nb}. 
For the higgs-portal model (where the DM annihilation proceeds through $p$-wave) 
this $\langle \sigma v \rangle$ is of the order of 
$10^{-31}$ $\rm cm^3 s^{-1}$ \cite{Coogan:2021rez, Coogan:2021sjs}. 

As we can see from fig.~\ref{fig:sv_mx_lim} (or \ref{fig:sv_mx_lim_future}), 
for the kinetic-mixing model, the combined indirect search 
bounds from existing observations already rule out 
the standard thermal relic sub-GeV DM scenario for $m_{\rm DM}$ up to 
$\sim800$ MeV. As the present work shows, 
the upcoming instrument {\sc Cosi} is expected to 
provide a further strong constraint on this parameter space. 
Note that there are mechanisms for producing the observed DM 
through different non-standard processes which can be considered 
to open up the parameter space for the sub-GeV DM candidates 
considered here; see for example \cite{Davoudiasl:2015vba, Evans:2019jcs, Asadi:2021bxp} 
for detailed discussions.  

For the higgs-portal model, 
as it can be seen from fig.~\ref{fig:sv_mx_lim} 
(or \ref{fig:sv_mx_lim_future}), the present indirect detection 
constraints on $\langle \sigma v \rangle$ 
are weaker compared to the standard thermal relic value. 
The upcoming {\sc Cosi} is expected to probe this thermal relic value 
only for $m_{\rm DM} \lesssim 2$ MeV. 
Note that, in this kind of scenario, the value of 
$\langle \sigma v \rangle$ that corresponds to the observed relic density 
can become significantly larger than the standard one shown here 
by considering a modified cosmological setup; see for example \cite{DEramo:2017gpl}.

We also note that, considering a Dirac fermion DM under the standard thermal scenario, 
Big Bang Nucleosynthesis (BBN) constrains the DM mass below a few MeV; 
see \cite{Depta:2019lbe, Sabti:2019mhn} and \cite{Fitzpatrick:2020vba}.

\subsubsection{CMB constraints}
In each $\langle \sigma v \rangle - m_{\rm DM}$ plane we show 
the constraint from Planck CMB observation~\cite{Planck:2018vyg} 
with a gray dash-dotted line. 
This is estimated for the two considered sub-GeV DM models 
(where DM is a Dirac fermion with particle-antiparticle symmetry) 
following \cite{Coogan:2019qpu} with the efficiency factors 
for $e^\pm$ and $\gamma$ from \cite{Slatyer:2015jla}. 
The CMB constraint for the kinetic-mixing model 
is comparatively much stronger (compared to the thermal line) 
since here the DM annihilation is $s$-wave. 
On the other hand, this constraint in the case of the higgs-portal model 
is much weaker since here the annihilation is $p$-wave. 
In this case, the DM velocity at recombination which is required 
to translate the CMB constraint on $\langle \sigma v \rangle$ 
to its present-day value is estimated considering an 
optimistic choice for the kinetic decoupling temperature 
of $x_{kd} = 10^{-6}$ (increasing which should relax the CMB constraint)~\cite{Essig:2013goa}. 
{Note here that, in the cases in which the vector portal model 
also features $p$-wave annihilations 
(e.g. if the DM is a complex scalar or a Majorana fermion), 
the CMB constraints are equally relaxed and 
the corresponding thermal relic lines go down. 
Instead the indirect detection limits obtained in our analysis are sensitive only to 
DM annihilation at the present time and therefore are quite independent of the $s$-wave/$p$-wave 
assumption if we consider, as usually done, a constant DM speed in the galactic halo.}

We see from fig.~\ref{fig:sv_mx_lim} (or \ref{fig:sv_mx_lim_future}) that, 
for the kinetic-mixing model, 
the indirect detection constraints 
derived using existing data are weaker compared to the bound from the CMB. 
Taking an optimistic approach using the standard 
astrophysical backgrounds for the photon observations,  
this comparison can be improved. 
For example, the optimistic bound from {\sc Integral} 
reaches at the same level of the CMB bound for 
$m_{\rm DM}$ in the ranges $\sim 100$ MeV - 1 GeV. 
On the other hand, as can be seen from fig.~\ref{fig:sv_mx_lim_future}, 
the upcoming {\sc Cosi} (with a year of observation towards the GC) 
can in principle improve the situation a lot by 
providing a good sensitivity to the secondary emissions 
produced from the DM annihilation under this model.
For example, it can probe values of $\langle \sigma v \rangle$ 
that lie almost an order of magnitude below the CMB bound 
for $m_{\rm DM} \gtrsim 100$ MeV. 
Considering the effects of a full galactic propagation of 
DM-induced $e^\pm$ on the corresponding 
secondary emissions, such {\sc Cosi} projections can be 
improved even further, especially below $m_{\rm DM} \simeq 70$ MeV, 
by reaching at the same level of the CMB bound 
or becoming stronger than this.

The situation is quite different in the case of the 
higgs-portal model. Fig.~\ref{fig:sv_mx_lim_future} (or \ref{fig:sv_mx_lim}) 
shows that, for this model, the combined indirect search bounds 
from $X$-ray, $\gamma$-ray and cosmic-ray observations 
(even the ones obtained with the conservative approach) are much stronger 
than the CMB bound for almost the entire DM mass range MeV-GeV, especially 
for $m_{\rm DM} \gtrsim 10$ MeV. 
The upcoming {\sc Cosi} (with a year of observation towards the GC) 
then will be able to probe for this model 
much deeper into the parameter space that lies several orders 
of magnitude below the CMB bound 
in the whole DM mass range 1 MeV -- 1 GeV. 

Note that the CMB constraint discussed above can be evaded independently 
for example in a scenario where the DM is partially produced 
from the decay of a heavier dark-sector particle 
after recombination; see \cite{DEramo:2018khz}. 
On the other hand, the indirect detection bounds or projections derived here 
remain unaltered in such a scenario. 

\subsubsection{Terrestrial constraints} 
\label{sec:Terrestrial_constraints}
For the kinetic-mixing model, we also compare our results 
with the combined exclusions and the future projections from 
different terrestrial experiments. 
These exclusions and projections, obtained from fig.~3 of \cite{Krnjaic:2022ozp} 
for the masses we are interested in, are shown 
in figs.~\ref{fig:sv_mx_lim}, \ref{fig:sv_mx_lim_future} 
and \ref{fig:sv_mx_lim_secondaries} 
by the dark and light gray shaded regions, respectively. 
For the projected sensitivities, we consider the 
future experiments that are starting their operations or have secured full funding. 
These terrestrial exclusions and projections are shown originally 
in the $y - m_{\rm DM}$ plane (with the quantity $y$ defined in eq.~\eqref{eq:sv_kineticMixing}), 
along with the contours corresponding to the observed relic density obtained 
under the thermal freeze-out scenario for different DM candidates 
(see also \cite{LDMX:2018cma, 10.1093/ptep/ptz106}).
Since the total pair-annihilation cross-section ($\langle \sigma v \rangle$) 
of DM into SM for a given $m_{\rm DM}$ is proportional to $y$ 
(as shown in eq.~\eqref{eq:sv_kineticMixing}), 
and considering that 
$\langle \sigma v \rangle_{\rm thermal} \simeq 10^{-26}$ $\rm cm^3 s^{-1}$ 
for the correct thermal relic density for DM, 
we convert the experimental constraints and projections from $y - m_{\rm DM}$ plane to 
the $\langle \sigma v \rangle - m_{\rm DM}$ plane using the relation 
$\langle \sigma v \rangle_{\rm exp} = \langle \sigma v \rangle_{\rm thermal} \, \times 
({y_{\rm exp}}/{y_{\rm thermal}})$. 
The quantity $y_{\rm thermal}$, indicating the thermal relic contour, 
corresponds to the case of a Majorana DM with a multiplication factor 
of 2 (since our DM is a Dirac fermion with particle-antiparticle symmetry). 
Note that these terrestrial constraints and projections are shown for a choice of the 
coupling parameter $\alpha_D = g^2_{V{\rm DM}} / 4\pi = 0.5$ 
used in most of the literatures. Assuming a smaller value of $\alpha_D$ 
will strengthen these combined constraints and projections; 
see, e.g., \cite{LDMX:2018cma} for a quantitative discussion. 
On the other hand, as mentioned in sec.~\ref{sec:Vector_model}, 
since the DM-induced photon and $e^\pm$ spectra in this model 
(for the masses we are interested) do not depend on $\alpha_D$, 
the indirect detection limits on $\langle \sigma v \rangle$ 
remain unaffected under the variation of this parameter. 

Figs. \ref{fig:sv_mx_lim} and \ref{fig:sv_mx_lim_future} show that, 
for the kinetic-mixing model, our combined indirect detection 
constraints (based on prompt and secondary photon signals) 
from existing $X$-ray + $\gamma$-ray observations as well as 
from {\sc Voyager-1} observation, obtained with the conservative approach 
(without assuming any background modeling), provide 
better exclusions than those from existing terrestrial experiments 
for a DM mass up to $\sim450$ MeV. 
In this mass range, these indirect detection bounds 
are stronger then the terrestrial ones by more than 1--2 orders 
of magnitude depending on $m_{\rm DM}$. 
Such constraints also exceed the projected sensitivities 
of upcoming terrestrial experiments 
for a DM mass up to $\sim200$ MeV.  
Taking an optimistic approach using the standard astrophysical backgrounds, 
such indirect detection bounds are improved by an order 
of magnitude and the above-mentioned mass ranges 
are extended up to $\sim800$ MeV and $\sim500$ MeV, respectively. 

From fig.~\ref{fig:sv_mx_lim_future} we see that 
the upcoming MeV telescope {\sc Cosi} will be able to explore a much 
larger region of the parameter space of the kinetic-mixing model, 
thanks to the significant leverage provided by 
its sensitivity to secondary emissions. 
With a year of observation towards the GC, 
{\sc Cosi} will be able to probe a vast parameter space 
that lies {\it well beyond} the reach of the present or 
future terrestrial experiments 
for a DM mass range that is extended up to $\sim600$ MeV. 
Hence it can complement these terrestrial experiments in the search for 
sub-GeV DM signals in this mass range.

\medskip

For the higgs-portal model with $m_S < m_{\rm DM}$, the DM pair-annihilation 
cross-section depends only on the coupling $g_{S{\rm DM}}$, 
but not on the mixing parameter $\sin \vartheta$ 
between the scalar mediator and the SM higgs.  
The only requirement is that there is some value of $\sin \vartheta$ 
consistent with the terrestrial observations, and that it is large 
enough that the decay length of the mediator is below the parsec scale \cite{Coogan:2021sjs}.
Hence, the complementarity between the indirect detection constraints and those from 
the terrestrial searches \cite{Krnjaic:2015mbs} can not be shown in the $\langle \sigma v \rangle - m_{\rm DM}$ plane. For this model, we compare our results with the 
cosmological constraints (CMB and relic abundance). 
In addition, like in \cite{Coogan:2021sjs}, we show (in fig.~\ref{fig:sv_mx_lim}) 
some contours of constant $g_{S{\rm DM}}$ to get an idea of 
reasonable values of $\langle \sigma v \rangle$. 
For example, the existing photon and cosmic-ray observations together are 
able to constrain the values of $g_{S{\rm DM}}$ between $\sim0.1$ -- a few $\times 10^{-3}$ 
for a DM mass up to a few hundreds of MeV. 
The near-future {\sc Cosi} will be able to improve this situation further.

{
\subsubsection{SN1987A constraints}
For the kinetic-mixing model, we also compare our DM indirect detection results 
with the Supernova 1987A (SN1987A) constraints on the same model from \cite{Chang:2018rso}. 
Considering the fiducial SN model and a choice for the coupling parameter 
$\alpha_D = 0.5$ (like in the case of the terrestrial constraints discussed above), 
we map the region of the parameter space excluded by SN1987A 
from the $y - m_{\rm DM}$ plane (presented in the right panel of figure 6 of \cite{Chang:2018rso}) 
into our $\langle \sigma v \rangle - m_{\rm DM}$ plane
(following the same prescription discussed above in sec.~\ref{sec:Terrestrial_constraints}). 
We find that this excludes a region 
$10^{-37} \, {\rm cm^3\,s^{-1}} \lesssim \langle \sigma v \rangle \lesssim 10^{-31} \, {\rm cm^3\,s^{-1}}$ 
(which falls below the level shown in fig.~\ref{fig:sv_mx_lim} or \ref{fig:sv_mx_lim_future}) 
for $m_{\rm DM}$ in the range 1 -- $\sim100$ MeV, 
leaving the other parts of the parameter space, in particular those to be probed by upcoming DM 
indirect detection searches, allowed.}

\section{Conclusions}
\label{sec:Conclusions}
In this work we perform a detailed study of the indirect searches of 
sub-GeV DM particles (in the MeV -- GeV mass range) in two representative 
realistic models, namely, the vector portal kinetic-mixing model and the higgs-portal model, 
based on the primary and secondary photon signals 
as well as the cosmic-ray signals produced in the Galaxy 
from the pair-annihilation of these DM particles. 
These models lead to the annihilation of DM particles into realistic 
final states including various light hadronic resonances. 
Regarding the photon observations, focusing on different observation regions 
centered around the GC, we include DM annihilation induced all possible 
secondary signals that result from Inverse Compton Scatterings (ICS), bremsstrahlung and 
In-flight positron annihilation (IfA). 
We adopt standard assumptions for the galactic environment 
and proper treatments for the galactic propagation of DM-induced $e^\pm$ that give rise 
to the secondary photons as well as the cosmic-ray signals. 
Based on the above-mentioned signals, we derive indirect detection constraints 
on the two considered sub-GeV DM models using existing $X$-ray/soft $\gamma$-ray data from 
experiments like {\sc Integral}, {\sc Comptel} and {\sc Fermi-Lat} and 
$e^\pm$ data from {\sc Voyager-1}. In parallel, we evaluate the projected sensitivity of 
the upcoming MeV telescope {\sc Cosi} (which is already selected to fly) 
in probing the parameter space of these DM models, based on all possible 
primary and secondary photon signals. 
We compare our indirect detection bounds and projections with other 
constraints/projections from cosmological observations and 
different terrestrial experiments. 

\medskip

We find that, thanks to the inclusion of secondary photons, our 
combined bound on the total DM annihilation rate $\langle \sigma v \rangle$ 
from existing $X$-ray and $\gamma$-ray observations 
turn out to be significantly stronger compared to the 
terrestrial constraints and future projections 
for a DM mass up to a few hundreds of MeV for the kinetic-mixing model. 
With a rather optimistic approach using the standard astrophysical backgrounds 
such bound on this model can reach at the level of the CMB constraint 
(which is expected to be the dominating constraint for this DM model) 
for $m_{\rm DM} \gtrsim 100$ MeV, based on mainly the ICS signal. 
For the higgs-portal model, on the other hand, 
this combined bound overcomes the CMB constraint 
by one to several orders of magnitude for $m_{\rm DM}$ above $\sim10$ MeV. 
{\sc Voyager-1} provides a better constraint for lower DM masses depending on the 
choice of the galactic propagation models. 

We show that the upcoming {\sc Cosi}, with a year of observation focusing towards the GC region, 
is going to provide a significant improvement to the 
above-mentioned constraints, thanks to its sensitivity towards the secondary photon emissions. 
For both the models, {\sc Cosi} can in principle probe a region of the parameter space that 
lies well beyond the reach of the existing indirect search or terrestrial experiments. 
For models like the kinetic-mixing case, where the CMB constraint 
is strong, {\sc Cosi} can probe values of $\langle \sigma v \rangle$ 
that are still well allowed by this constraint for a DM mass 
above $\sim100$ MeV (based on mainly the ICS signal) 
as well as below $\sim10$ MeV (based on mainly bremsstrahlung, IfA and prompt photon signals) 
depending on the choice for the Galactic propagation. 

\medskip

To summarize, the present work highlights in detail the significance of leveraging 
the galactic DM annihilation induced secondary emissions to 
evaluate the indirect detection bounds on realistic sub-GeV DM models 
as well as the corresponding sensitivity projections from 
upcoming MeV telescopes like {\sc Cosi}.

\small
{\subsubsection*{Acknowledgments}
\footnotesize{
We thank Peter Reimitz for useful discussions and collaboration in the initial stages of this work. 
We thank Pierre Salati for providing us with the {\sc Voyager}-1 data. 
We also thank Lukas Simon and Andrea Caputo for useful discussions. 
M.C and A.K. acknowledge the hospitality of the Institut d'Astrophysique de Paris ({\sc Iap}) where part of this work was done. M.C. also acknowledges the hospitality of the Flatiron Institute (Simons Foundation) and of the Physics Department of New York University. 

\noindent Funding and research infrastructure acknowledgments: 
Research grant {\sl DaCoSMiG} from the {\sc 4eu+} Alliance (including Sorbonne Université). Support from the Institut Henri Poincaré ({\sc Uar} 839 {\sc Cnrs}-Sorbonne Université), and the LabEx {\sl CARMIN} (ANR-10-LABX-59-01).
}}

\appendix

\section{Power emitted in different secondary processes}
\label{sec:Powers_secondary}

Here we describe the powers emitted in photons 
through different secondary processes considered in this work 
for the DM annihilation in the Galaxy. 

\begin{itemize}

\item \underline{\bf Inverse Compton Scatterings (ICS):} 
The differential power emitted due to the IC scatterings of an $e^\pm$  
on each ambient photon bath $i$ in the interstellar radiation fields (ISRFs) is expressed as 
(see \cite{Cirelli:2009vg, Colafrancesco:2005ji, Cirelli:2024ssz}),
\begin{equation}
\mathcal{P}^i_{\rm ICS} (E_\gamma, E_e, \Vec{x}) = c \, E_\gamma \int d\epsilon \, n^{\rm ISRF}_i (\epsilon, \Vec{x}) \,\, 
\sigma_{\rm IC} (\epsilon, E_\gamma, E_e) \, ,
\label{eq:P_ICS}
\end{equation}
where $i\rightarrow$ CMB, infrared (IR) or starlight (SL) photon field, 
with the differential photon number density $n^{\rm ISRF}_i (\epsilon, \Vec{x})$ 
(which is the same one used in \cite{Buch:2015iya, Cirelli:2020bpc, Cirelli:2023tnx, Cirelli:2025qxx}). 
The quantity $\sigma_{\rm IC}$ denotes the Klein-Nishina cross-section, 
obtained from \cite{Cirelli:2009vg, Colafrancesco:2005ji, Cirelli:2024ssz}. 
The limit of integration over the ambient photon energy $\epsilon$ 
in eq.~\eqref{eq:P_ICS} is determined by the kinematics of the IC scattering 
$1 \le q \le m^2_e/4E^2_e$, where 
$q \equiv \frac{E_\gamma m^2_e}{4 \epsilon E_e (E_e-E_\gamma)}$. 
A graphical illustration of $P^i_{\rm ICS}$ (corresponding to the three ISRF components) 
as a function of the produced photon energy $E_\gamma$ (for different input $e^\pm$ energies) 
can be found in \cite{Cirelli:2020bpc}. The total power in the ICS process is obtained as: 
$\mathcal{P}_{\rm ICS} = \sum_{i\in {\rm ISRF}} \mathcal{P}^i_{\rm ICS}$.

\item \underline{\bf Bremsstrahlung:} 
The bremsstrahlung power emitted due to the scattering of an $e^\pm$ 
(with $E_e > E_\gamma$) with the gas particles is given by (see \cite{Buch:2015iya}): 
\begin{equation}
\mathcal{P}_{\rm brem} (E_\gamma, E_e, \Vec{x}) = c \, E_\gamma \sum_{i} 
n_i(\Vec{x}) \, \frac{d\sigma^{\rm brem}_i}{dE_\gamma} (E_e, E_\gamma) \, .
\end{equation}
Here $n_i(\Vec{x})$ describes the number density distribution of each of the 
gas species (ionic, atomic and molecular) and 
$\frac{d\sigma^{\rm brem}_i}{dE_\gamma} (E_e, E_\gamma)$ is the 
corresponding differential scattering cross-section for bremsstrahlung. 
The gas maps used here are the same ones used in 
\cite{Buch:2015iya, Cirelli:2020bpc, Cirelli:2023tnx, Cirelli:2025qxx}. 

\item \underline{\bf In-flight annihilation (IfA):} 
The power emitted due to the in-flight annihilation of an $e^+$ can be expressed as \cite{Bartels:2017dpb}: 
\begin{equation}
\mathcal{P}_{\rm IfA} (E_\gamma, E_e, \Vec{x}) = c \, E_\gamma \, n_{e^-}(\Vec{x}) \, \beta_e \, 
\frac{d\sigma^{\rm IfA}}{dE_\gamma} (E_e, E_\gamma) \, .
\end{equation}
Here $\frac{d\sigma^{\rm IfA}}{dE_\gamma} (E_e, E_\gamma)$ 
is the differential cross-section related to the IfA process (see \cite{Bartels:2017dpb}), 
$\beta_e$ is the dimensionless velocity of the $e^+$ and 
$n_{e^-}(\Vec{x})$ is the number density of target electrons in the galactic medium 
(obtained from \cite{Buch:2015iya}). 
The kinematics for the process is 
$\gamma_e (1-\beta_e) \leq 2 E_\gamma/m_e - 1 \leq \gamma_e (1+\beta_e)$, 
with $\gamma_e = E_e/m_e$.

\end{itemize}

\section{Dependence on the DM profile}
\label{sec:DM_profiles} 
\begin{figure*}[!t]
\centering
\hspace{-2mm}
\includegraphics[width=0.5\textwidth]{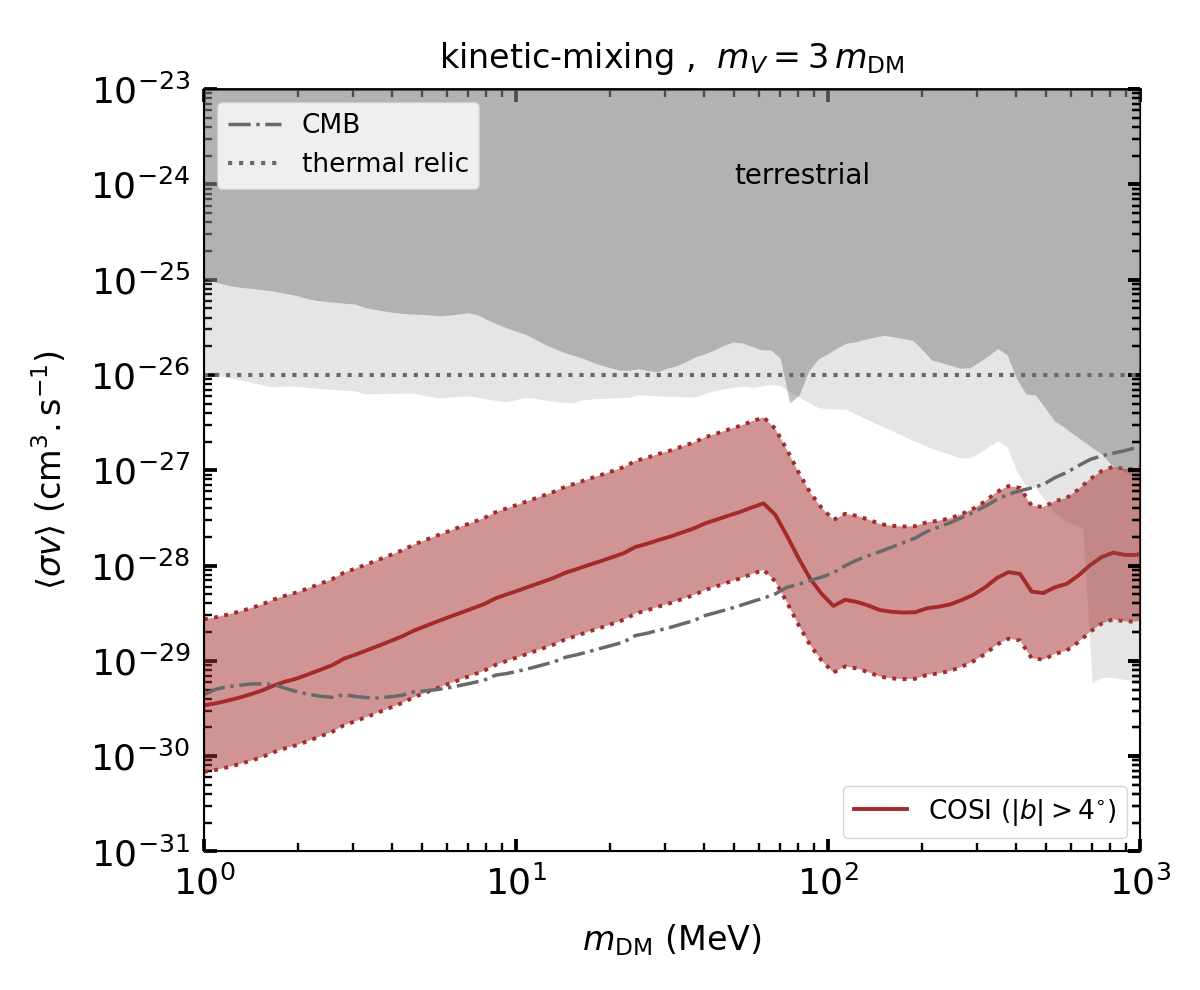}
\hspace{-2mm}
\includegraphics[width=0.5\textwidth]{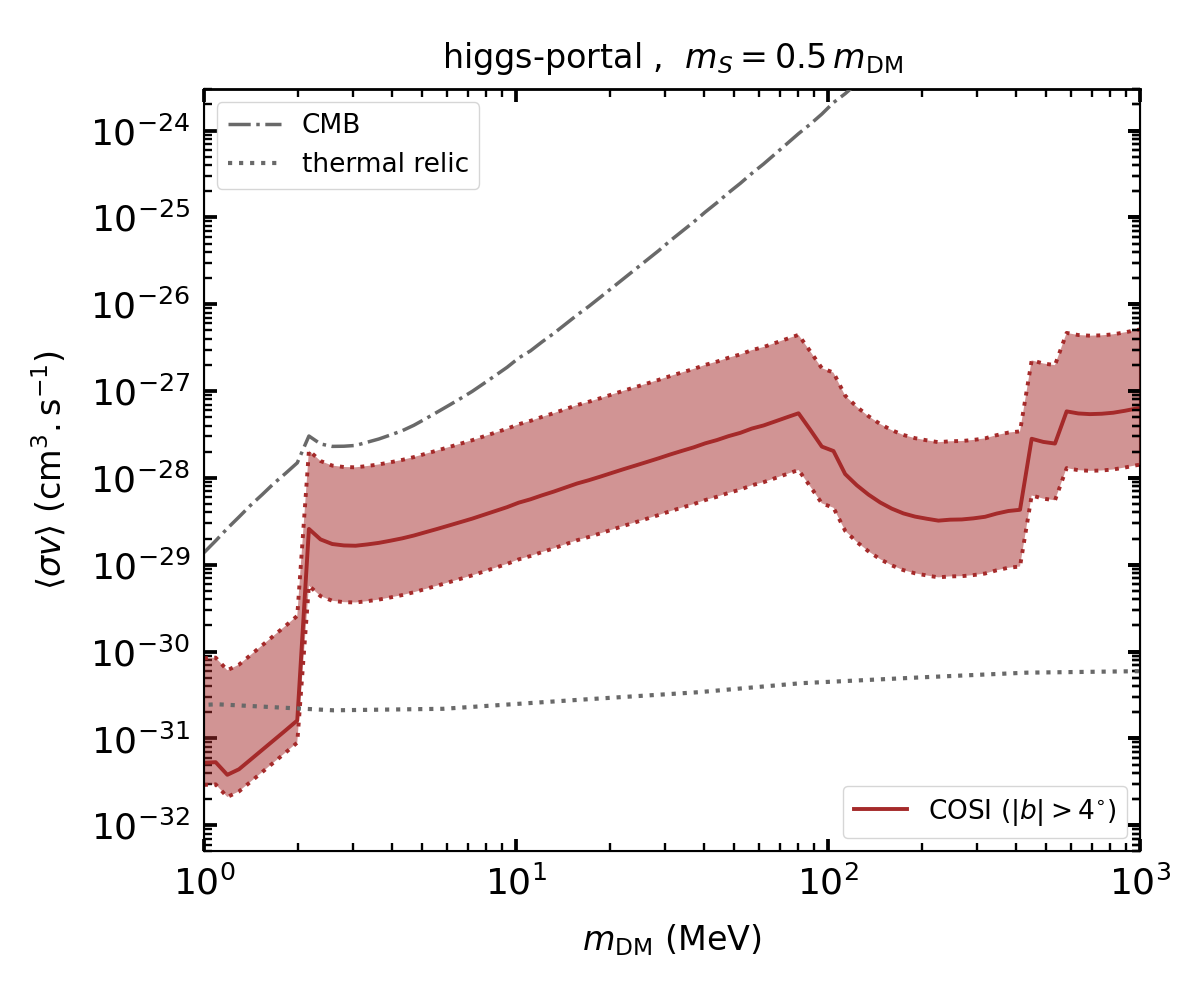}\\
\vspace{-3mm}
\caption{\em Effect of {\bfseries varying the DM profile}. 
Brown solid line: the {\sc Cosi} projection obtained considering 
the NFW DM profile of eq.~\ref{eq:rho_profile} (the same as the one shown in fig.~\ref{fig:sv_mx_lim_future}). 
Upper edge of the band: corresponding to the same Isothermal (cored) 
DM profile discussed in \cite{Cirelli:2025qxx}. 
Lower edge of the band: corresponding to the Einasto profile from 
\cite{2019JCAP...10..037D} (for the baryonic model B2) 
with the profile parameters adjusted within their $1\sigma$ uncertainty 
to maximize the $J$-factor (like in \cite{Coogan:2021sjs}).}
\label{fig:sv_mx_lim_DM_profiles}
\end{figure*}

In fig.~\ref{fig:sv_mx_lim_DM_profiles} we illustrate the 
effect of considering different DM profiles on the projected sensitivities of {\sc Cosi}, for 
the two considered DM models. In each case, the brown solid line corresponds to 
the projection obtained with the NFW profile of eq.~\ref{eq:rho_profile} and 
is same as the one shown in fig.~\ref{fig:sv_mx_lim_future}. The associated brown band 
indicates the variation due to different choices for the DM profile. 
The upper edge of the band corresponds to the truncated Isothermal (cored) 
profile {$\rho^{\rm Iso}_{\rm DM}(r) = {\rho_0} / \left(1 + \left(\frac{r}{r_s}\right)^2\right)$, 
with the corresponding parameters $\rho_0 = 2.112 \, {\rm GeV\,cm^{-3}}$ and $r_s = 4$ kpc} 
tabulated in \cite{Cirelli:2024ssz} (also used in \cite{Cirelli:2025qxx}). 
On the other hand, the lower edge of the band 
corresponds to the Einasto profile 
{$\rho^{\rm Ein}_{\rm DM}(r) = \rho_0 \, {\rm exp} \left\lbrace -\frac{2}{\alpha} 
\left(\left(\frac{r}{r_s}\right)^\alpha - 1\right) \right\rbrace$} 
from \cite{2019JCAP...10..037D} (corresponding the baryonic model B2) with the profile parameters 
adjusted within their $1\sigma$ uncertainty to maximize the $J$-factor (like in \cite{Coogan:2021sjs}). 
{This corresponds to 
$\alpha \simeq 0.1$, $r_s \simeq 6.5$ kpc and $\rho_{\rm DM}(r_\odot) \simeq 0.4 \, {\rm GeV\,cm^{-3}}$}.
Fig.~\ref{fig:sv_mx_lim_DM_profiles} shows that, depending on the choice of DM profile, 
the indirect detection constraints on the DM models 
presented in fig.~\ref{fig:sv_mx_lim_future} can vary by a factor of a few in both directions.


\bibliographystyle{JHEP}
\bibliography{bibliography.bib}

\end{document}